\documentclass[12pt, draftclsnofoot, onecolumn]{IEEEtran}
\usepackage{caption}
\usepackage{floatrow}
\newfloatcommand{capbtabbox}{table}[][\FBwidth]
\usepackage{blindtext}
\usepackage[font=scriptsize]{caption}
\usepackage{cite,graphicx,amsmath,amssymb}
\usepackage{graphicx,caption}
\usepackage{amsfonts,amssymb}
\usepackage{algorithm,algcompatible,algorithmicx,amsmath}
\usepackage{subfigure}
\usepackage{fancyhdr}
\usepackage{balance}
\usepackage{xcolor}
\usepackage{bm}
\usepackage{amsthm}
\usepackage{multirow}
\usepackage{flafter}
\usepackage{mathrsfs}
\usepackage{url}
\usepackage{enumerate}
\usepackage{stfloats}
\usepackage{setspace}
\providecommand{\keywords}[1]{\textbf{\textit{Index terms---}} #1}
\theoremstyle{definition}
\newtheorem{definition}{Definition}
\newtheorem{remark}{Remark}
\newtheorem{theorem}{Theorem}
\begin{document}

\title{IRS Assisted NOMA Aided Mobile Edge Computing with Queue Stability: Heterogeneous Multi-Agent Reinforcement Learning}
\author{Jiadong~Yu,\IEEEmembership{ Member,~IEEE,} Yang~Li, Xiaolan~Liu,\IEEEmembership{ Member,~IEEE,}\\
Bo~Sun,\IEEEmembership{ Member,~IEEE,}
Yuan~Wu,\IEEEmembership{ Senior Member,~IEEE,}\\
Danny~H.K.~Tsang,\IEEEmembership{ Fellow,~IEEE} %~\IEEEmembership{Student Member,~IEEE,}
%~Yue~Gao
\thanks{\protect\onehalfspacing J. Yu is with the Internet of Things Thrust, The Hong Kong University of Science and Technology (Guangzhou), Guangzhou, Guangdong 511400, China (Email: jiadongyu@ust.hk).}
\thanks{\protect\onehalfspacing Y. Li is with the State Key Laboratory of Internet of Things for Smart City and Department of Computer and Information Science, University of Macau, Macao SAR, China (Email: yb87469@um.edu.mo).}
\thanks{\protect\onehalfspacing X. Liu is with the Institute for Digital Technologies, Loughborough University, London E20 3BS, U.K. (Email: xiaolan.liu@lboro.ac.uk).}
\thanks{\protect\onehalfspacing B. Sun is with the Department of Electronic and Computer Engineering, The Hong Kong University of Science and Technology, Clear Water Bay, Hong Kong SAR, China (Email: bsunaa@connect.ust.hk).}
\thanks{\protect\onehalfspacing Y. Wu (Corresponding author) is with The State Key Lab of Internet of Things for Smart City, and also with the Department of Computer and Information Science, The University of Macau, Macao SAR, China (Email: yuanwu@um.edu.mo).}
\thanks{\protect\onehalfspacing D.H.K. Tsang is with the Internet of Things Thrust, The Hong Kong University of Science and Technology (Guangzhou), Guangzhou, Guangdong 511400, China, and also with the Department of Electronic and Computer Engineering, The Hong Kong University of Science and Technology, Clear Water Bay, Hong Kong SAR, China (Email: eetsang@ust.hk).
}
}
\maketitle

\begin{abstract}
By employing powerful edge servers for data processing, mobile edge computing (MEC) has been recognized as a promising technology to support emerging computation-intensive applications. Besides, non-orthogonal multiple access (NOMA)-aided MEC system can further enhance the spectral-efficiency with massive tasks offloading. However, with more dynamic devices brought online and the uncontrollable stochastic channel environment, it is even desirable to deploy appealing technique, i.e., intelligent reflecting surfaces (IRS), in the MEC system to flexibly tune the communication environment and improve the system energy efficiency. In this paper, we investigate the joint offloading, communication and computation resource allocation for IRS-assisted NOMA MEC system. We firstly formulate a mixed integer energy efficiency maximization problem with system queue stability constraint. We then propose the Lyapunov-function-based Mixed Integer Deep Deterministic Policy Gradient (LMIDDPG) algorithm which is based on the centralized reinforcement learning (RL) framework. To be specific, we design the mixed integer action space mapping which contains both continuous mapping and integer mapping. Moreover, the award function is defined as the upper-bound of the Lyapunov drift-plus-penalty function. To enable end devices (EDs) to choose actions independently at the execution stage, we further propose the Heterogeneous Multi-agent LMIDDPG (HMA-LMIDDPG) algorithm based on distributed RL framework with homogeneous EDs and heterogeneous base station (BS) as heterogeneous multi-agent. Numerical results show that our proposed algorithms can achieve superior energy efficiency performance to the benchmark algorithms while maintaining the queue stability. Specially, the distributed structure HMA-LMIDDPG can acquire more energy efficiency gain than centralized structure LMIDDPG.
\end{abstract}


\section{Introduction}
%device increase
%mec exist
With the explosive growth of online devices, the Internet of Things (IoT) era brings the innovation applications, such as smart home, intelligent transportation, industrial automation, and smart healthcare\cite{7123563}. These emerging data-driven and computation-intensive application services lead to more stringent requirements for system performance such as low latency, low energy consumption, and privacy preserving, which greatly stimulate the rapid development of wireless communication technology. In recent years, mobile edge computing (MEC), which deploys edge servers at base stations (BSs) to extend the cloud-computation capabilities, has been recognized as a promising technology to tackle the long latency backhaul-limitation and computation resource-demanding challenges faced by cloud computing\cite{7488250}.

To meet the massive connectivity demand and improve the spectral efficiency, non-orthogonal multiple access (NOMA), utilizing superposition coding and successive interference cancellation (SIC) techniques, has been recognized as an essential communication technique to support MEC for large capacity and high data rates\cite{9113305}. Conventionally, in NOMA-aided systems, the communication channels are highly stochastic and cannot be tuned. Therefore, another appealing technique called intelligent reflecting surfaces (IRS) has drawn unprecedented attention in wireless communications. The key advantage of IRS is that the reflection elements can be dynamically tuned with specific phases and amplitudes so that they can collaboratively forward the impinging waves towards target directions\cite{9240028}. It has been demonstrated that the IRS-assisted NOMA-aided system can enable flexible control on the communication channel gains and enhance system energy efficiency\cite{9241881}. However, to our best knowledge, due to the extreme complex joint resource allocation problems, the study of IRS-assisted NOMA-aided MEC is still in its infancy.
\subsection{Related Work}
%RL has been widely used to solve the MEC work

The execution of MEC requires the users to offload the computation tasks to the edge server through wireless communications. To meet various real-world demands (i.e., low latency, low energy consumption, high energy efficiency), jointly making task offloading decisions and allocating the communication resources have always been challenging\cite{8016573}. Due to the highly dynamic characteristics of the MEC systems, the system modeling cannot fully achieve real-time accuracy. Thus, using traditional optimization methods for solution design may result in a loss of system performance. In recent years, data-driven methods represented by Reinforcement Learning (RL) have gained increasingly attention. RL is a model-free approach and does not require real-time and complete prior information of the edge system, and thus it is considered to be an emerging effective solution for task scheduling and resource allocation in highly dynamic MEC networks\cite{8972358}. According to different RL based resource allocation algorithm schemes, the research on MEC resource allocation of multi-user systems can be divided into two categories: centralized resource allocation\cite{8972358,8598893,8493155} and distributed resource allocation\cite{9205989,9485089}. 
%7762913,7956189,/8789642,9351538,
%centralized:traidition

In centralized resource allocation management scheme, the edge computing server obtains all mobile information, including channel status information, computing requests, and so on, and then applies different RL to carry out resource allocation strategy designs. Then the central BS informs the optimum strategy to mobile devices for execution. 
%In \cite{7762913}, the authors formulated the optimal resource allocation into a convex problem by minimizing the weighted mobile energy consumption. Then the optimal policy can be executed based on the derived offloading priority function. In \cite{7956189}, the Lyapunov optimization based low-complexity online algorithm has been proposed to minimize the long-term average weighted sum power consumption of both mobile devices and edge servers. 
In \cite{8598893}, a deep reinforcement learning (DRL) based offloading scheme for an IoT device with energy harvesting to select the edge device and the offloading rate has been proposed. Moreover, based on the task queue state, energy queue state and channel quality, the more advanced RL algorithm called double deep Q network strategy has been proposed to learn the optimal task offloading and energy allocation to maximize the system long-term utility\cite{8493155}. {However, due to the centralized information interaction and strategy broadcasting, the aforementioned centralized schemes will cause excessive communication overheads and information security concerns.}

To alleviate the heavy communication overheads caused by centralized scheme, distributed resource allocation frameworks have been investigated. 
%There are different distributed research methods on MEC, such as game theory\cite{8789642} and decentralized consensus optimization\cite{9351538}. In \cite{8789642}, the joint computation offloading and coin-loaning for blockchain-empowered MEC problem has been studied. To prove the existence of a pure-strategy Nash equilibrium, a distributed algorithm has been designed. In \cite{9351538}, a class of mini batch stochastic alternating direction method of multipliers algorithms to address the communication bottleneck and the slow response of the end devices was proposed.
The most typical of which is multi-agent reinforcement learning (MARL) \cite{9372298}. The MARL has the following advantages: firstly, agents learn their strategies in the distributed manner by observing local environment. Secondly, agents can be designed to share experience with each other based on customized purposes. Thirdly, multi-agent is more robust when some agents fail, as the remaining agents can take over the tasks. In \cite{9205989}, the computation offloading mechanism with resource allocation has been formulated as a stochastic game. The independent learners based multi-agent Q-learning algorithm has been proposed. In addition, the authors designed a distributed multi-agent deep reinforcement learning scheme to minimize the overall energy consumption of in the small cell networks\cite{9485089}. To further decrease the computation complexity and communication overheads of the training process, a federated DRL scheme which only share the model parameters has also been proposed in \cite{9485089}. 

Energy-efficient and spectral-efficient task offloading is the key to the success of MEC, which drives the deployment of NOMA in MEC offloading \cite{7973146}. Explicitly, NOMA allows multiple wireless devices to occupy the same frequency and time slot resources to transmit data to the BS, and BS uses SIC technology to decode the received information. Based on the system optimization metrics, the existing NOMA-aided MEC works are generally divided into two categories: energy consumption\cite{9679390,9393794} 
%\cite{8267072,\,8972932
and time delay\cite{8794550,9340353}. %To minimize the system energy consumption, the authours \cite{8267072} designed an efficient heuristic algorithm for user clustering and computing resources allocation, and further formulated a convex optimization problem for power control in each NOMA cluster. 
In \cite{9679390}, a hybrid NOMA-MEC offloading strategy, which combines conventional orthogonal multiple access and pure NOMA was proposed. Moreover, the multi-objective optimization problem has been formulated to minimize the energy consumption. Similarly, the authors also proposed an energy consumption optimization problem by optimizing the resource allocation and sub-channel assignment with the latency constraint in NOMA enabled MEC in \cite{9393794}. To minimize the time delay in the NOMA-MEC system, the authors in \cite{8794550} discussed that under the same power constraint, the deployment of NOMA can achieve substantial time delay reduction compare to time division multiple access (TDMA). 
%In \cite{8972932}, the differentiated offloading delay and co-channel interference has been characterized for a pair of NOMA users. 
To reduce the user average offloading delay, the optimal power allocation is obtained by convex programming, followed by iteratively semidefinite relaxation and convex-concave for NOMA user paring and offloading decision making. To minimize the maximal offloading latency of NOMA users, the authors exploited the successive convex approximation method to solve the complicated non-convex problem\cite{9340353}. Apart from aforementioned research, there are several works investigated the RL framework for NOMA-MEC\cite{9113721,9467317}. An online algorithm based on DRL framework to learn the near-optimal solution by minimizing the total energy consumption of end devices in multitask NOMA-MEC system was proposed in \cite{9113721}. Additionally, to minimize the long-term network computation cost, in \cite{9467317}, the authors proposed a cooperative multi-agent deep reinforcement learning framework to learn the decentralized policies at end users for task offloading decisions and local execution power allocation.

Recently, as a revolutionizing technology, IRS has been proved to be able to improve the performance of wireless communications with new degrees of freedom through highly controllable intelligent signal reflection \cite{8910627}. A few works exploited the beneficial role of IRSs in MEC, and they can be divided into two categories based on different optimization metrics: energy consumption \cite{9388935} and system throughput\cite{9270605, 9516969}. Authors in \cite{9388935} designed the resource allocation problem in the wireless powered IRS-MEC system by utilizing the alternating optimization technique and successive convex approximation method to minimize the system energy consumption. Differently, \cite{9270605} focused on optimizing IRS-MEC system's sum computational bits and the Lagrange dual method and Karush-Kuhn-Tucker conditions were proposed.  What's more, \cite{9516969} also focused on maximizing the total computation bits maximization problem for IRS-enhanced wireless powered MEC networks and the iterative algorithm was proposed to solve the non-convex non-linear optimization problem.
%the block coordinate descending (BCD) based algorithms aimed at maximizing the total completed task-input bits of all users in IRS-MEC system were proposed \cite{9380744}.
%What's more, \cite{9270605} also focused on optimizing IRS-MEC system's sum computational bits and the Lagrange dual method and Karush-Kuhn-Tucker conditions were proposed. 

\subsection{Motivation and Contributions}
%optimization metrics
%solving algorithms.
%system model
As discussed above, with the development of wireless communications, the research in MEC has evolved from fundamental communication environment \cite{8972358,8598893,8493155,9205989,9485089} to NOMA-aided 
\cite{9679390,9393794,8794550,9340353} and IRS-assisted systems\cite{9388935,9270605,9516969}. Despite the advantage of NOMA-aided MEC system with resource-efficient transmission, there are several limitations, such as the fluctuation of wireless channels, path loss, high user mobility, and so on that can influence the performance of the offloading communication. Although increasing the number of antennas at both transmitter and receiver can overcome such problems, the higher energy consumption occurs at the same time. As an near-zero energy consumption technique, IRS has been identified to support NOMA communication with enhanced performance from four aspects: tuned channel gains, improved fair power allocation, enhanced coverage range, and high energy efficiency\cite{9241881}, which motivates us to explore the IRS-assisted NOMA-aided MEC system.
%Therefore, the IRS-assisted NOMA-aided communication is considered in this paper.9679390,9393794
For optimization metrics design, minimizing energy consumption\cite{9679390,9393794} or time delay\cite{8794550,9340353}
%\cite{8794550,8972932,9340353}
are two widely explored problems in NOMA-enable MEC. Apart from saving energy consumption\cite{9388935}, system computation throughput\cite{9270605, 9516969} has been considered for IRS-assisted MEC. However, it is necessary to design a customized optimization metric for IRS-assisted NOMA-aided MEC system that can include both energy consumption, system throughput and time latency at the same time. 

Although the conventional optimization algorithms (i.e., Block coordinate descent (BCD), successive convex approximation, and so on) have been broadly proposed to solve different multi-convex or non-convex resource allocation problems in MEC, due to the computational difficulty and the requirements of the statistical knowledge of the MEC systems, it is impractical to deploy them in the highly dynamic environment. Variant RL frameworks become one of the efficient alternatives to solve the complex joint resource allocation problems without the prior statistical knowledge\cite{8972358,8598893,8493155,9205989,9485089,9467317,9449944}. Normally, the resource allocation tasks in MEC are mixed integer problems. However, most of the existing RL-based algorithms are limited to the action space with pure discrete 
\cite{8972358,8598893,8493155,9205989,9485089} or pure continuous\cite{9467317} action in MEC systems, which motivates us to consider the mixed integer mapping in the algorithm design. Although \cite{9449944} considered the multi-stage stochastic mixed integer non-linear programming problem and applied the centralized structure of RL framework to solve the decoupled problems, it is inevitable to consider the distributed manner MARL framework which allows end users to make decisions without outside information and can further save the heavy communication overheads. Moreover, the existing MARL frameworks on solving the MEC allocation problems mainly have homogeneous agents which have similar state and action spaces. However, there is no work considering the agents with different state and action spaces.

Motivated by the aforementioned literature review, we consider the IRS-assisted NOMA-aided MEC system and propose the energy efficiency with queue stability as the evaluation metric for optimal resource allocation strategy design. We further design both centralized RL framework and distributed MARL framework to solve the problem with mixed integer action space. By considering the different characteristics of the devices in MEC, the heterogeneous agents are designed in MARL framework. The main contributions are summarized as follows.
\begin{itemize}
    \item We formulate the energy efficiency maximization problem with queue stability in IRS-assisted NOMA-aided MEC system by optimizing the mixed integer action space, which consists of edge computing decision, local computing decision, transmit power allocation, local computing frequency, edge computing frequency and IRS phase shifts. 
    \item We further derive the upper-bound of the Lyapunov drift-plus-penalty function based on our proposed problem, which can reflect the system throughput, system power consumption and the queue stability of the designed problem. 
    \item The centralized RL framework algorithm named as Lyapunov-function-based Mixed Integer Deep Deterministic Policy Gradient (LMIDDPG) is proposed. Specifically, the proposed framework can map the continuous space into the mixed integer action space and the reward function is designed based on the derived Lyapunov upper-bound. 
    \item To save the communication interaction overheads and preserve privacy at the execution stage, the distributed RL framework algorithm named as Heterogeneous Multi-agent LMIDDPG (HMA-LMIDDPG) is proposed. Specially, there are heterogeneous multiple agents which include homogeneous end devices (EDs) and heterogeneous BS. 
    \item The proposed LMIDDPG and HMA-LMIDDPG can maintain the queue stability and outperform benchmark  algorithms with higher energy efficiency. Moreover, the distributed framework HMA-LMIDDPG can achieve further performance gain compared to the centralized framework LMIDDPG.
\end{itemize}

The rest of this paper is organized as follows. The IRS-assisted NOMA-aided MEC system model is introduced in Section~\ref{section2}. In Section~\ref{section3}, the task offloading and resource allocation problem is formulated. The proposed algorithm LMIDDPG and HMA-LMIDDPG are presented in Section~\ref{section4}. Numerical results and conclusions are then provided in Section~\ref{section5} and Section~\ref{section6}, separately.
%\textit{Notation:} Throughout the paper, we use the following common notation. The complex numbers are denoted by $\mathbb{C}$. The transpose and the conjugate transpose are denoted by $ (\cdot) ^\text{T}$, $(\cdot) ^\text{H}$ %$(\cdot)^{\dag}$ respectively. $\boldsymbol{I}$ is the identity matrix. $ \mathcal{C}\mathcal{N}$ (mean, covariance) indicates a complex Gaussian random vector with defined mean and covariance.
\section{System Model}
\label{section2}
\begin{figure*}
\begin{floatrow}
\ffigbox{%
 \includegraphics[width=7cm]{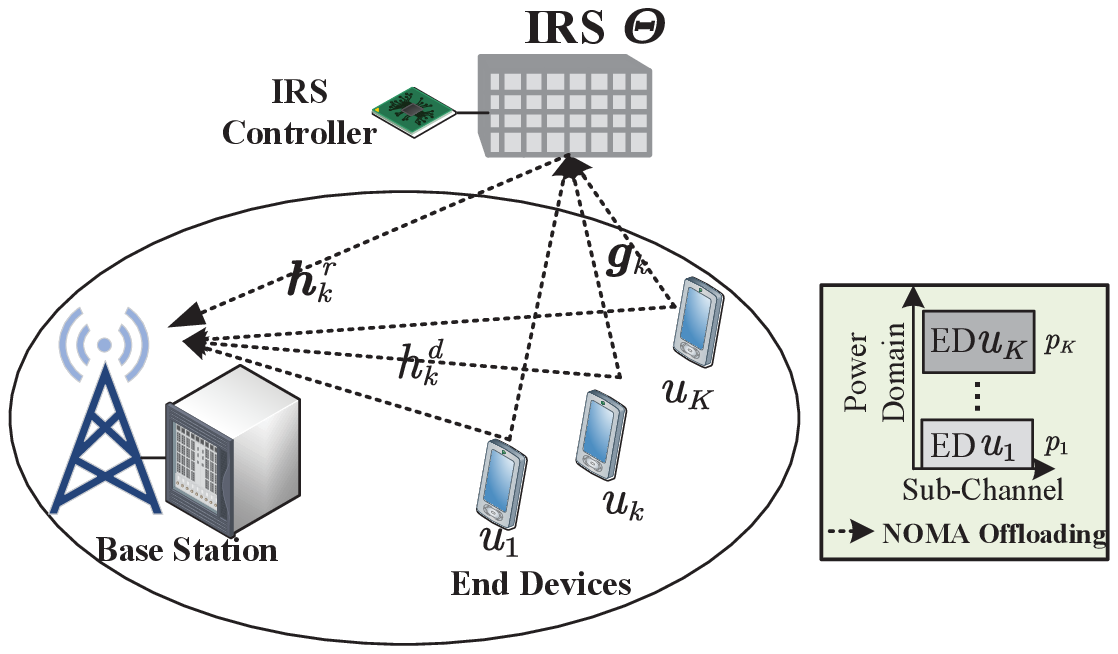}\\%
}{%
  \captionof{figure}{IRS-assisted NOMA-aided MEC system.}\label{system model}%
}
\ffigbox{%
  \includegraphics[width=8cm]{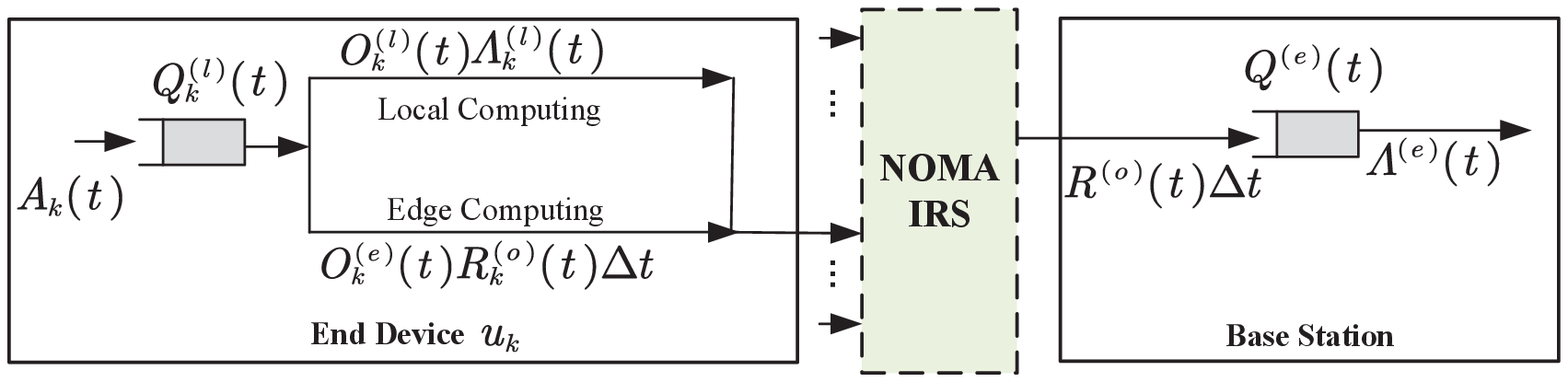}\\%
}{%
  \captionof{figure}{Task queuing model at ED $u_k$ ED and BS.}\label{task}%
}
\end{floatrow}
\end{figure*}
As illustrated in Fig. \ref{system model}, we consider the IRS-assisted NOMA-aided MEC system, where a set of single antenna devices, denoted by $\mathcal{U}=\{u_1,...,u_{k},...,u_{K}\}$ with $k\in\mathcal{K}=\{1,2,...,K\}$ are served by one BS and one IRS. The EDs can be different kinds of sensors and mobile devices that are energy-constrained and computation-limited. The system is assumed to operate at a time slotted structure which is indexed by an integer $t\in\mathcal{T}=\{1,2,...,T\}$, and the slot length is $\Delta T$. At each time slot $t$, the computation task is assumed to be generated by ED $u_k$ as the independent and identically distributed sequence of Bernoulli random variables with common parameter $\zeta_{k}(t)\in [0,1]$, and the task generation indicator $\mathcal{G}_{k}(t)=\{0,1\}$ (i.e., $\zeta_{k}(t)= \mathbb{P} (\mathcal{G}_{k}(t)=1)=1-\mathbb{P} (\mathcal{G}_{k}(t)=0)$). Specifically, $\mathcal{G}_{k}(t)=0$ indicates that there is no task generated by ED $u_k$ at time slot $t$. Moreover, $\mathcal{G}_{k}(t)=1$ represents that a task with constant size $A_{k}(t)$ in bits is generated. Since the BS which has much higher computation ability, the intensive tasks generated at EDs can either be processed locally or offloaded to the BS through NOMA. In this work, we consider the partial offloading\footnote{To be noticed, our work can be easily converted to binary offloading, in which each ED $u_{k}$ makes its own decisions on whether to offload the task (i.e., ${O}=\{1,0\}$). When ${O}_{k}(t)={1}$, the generated task is offloaded to the BS for edge computing. ${O}_{k}(t)={0}$ represents the task is executed locally at ED $u_k$.}, which means each ED makes its own decision on whether to offload the task (i.e., ${O_{k}^{(e)}}=\{1,0\}$) and/or process locally (i.e., ${O}_{k}^{(l)}=\{1,0\}$). When ${O}_{k}^{(e)}(t)={1}$, the data from the queue is offloaded to the BS for edge computing. ${O}_{k}^{(l)}(t)={1}$ represents the the data from the queue is executed locally at ED $u_k$. Specifically, EDs can process the data from the queue both locally and offloaded simultaneously (i.e., ${O}_{k}^{(l)}(t)={1}, {O}_{k}^{(e)}(t)={1}$).

\subsection{Communication Model}
The overall bandwidth $B$ is equally divided into $N$ channels, 
%and they can be denoted by $\mathcal{N}=\{1,2,...,N\}$,
each with the bandwidth as $B^{'}=\frac{B}{N}$. 
%For NOMA-based IRS-aided system, we assume that there are ${K}$ users occupying the $n^{th}$ channel and each ED can only be assigned to one channel. 
In this work, we focus on exploring the sub-channel NOMA-based IRS-aided system. To be specific, we assume that there are ${K}$ users occupying the same sub-channel. The movement of EDs are assumed to follow Gauss-Markov mobility model\cite{7080887}. To be specific, the velocity of $u_k$ at time slot $t$ can be modeled as $v_{k}(t)=\alpha\cdot v_{k}(t-1)+(1-\alpha)\cdot\bar{v}+\bar{\kappa}\cdot\sqrt{1-\alpha^2}\cdot W_{k}(t-1),$ where $v_{k}(t)=\left[v_{k}^{x}(t),v_{k}^{y}(t)\right]$ is the velocity vector, $W_{k}(t)=\left[w_{k}^{x}(t),w_{k}^{y}(t)\right]\sim\mathcal{N}(0,\varsigma^2)$ denotes the uncorrelated random Gaussian process, and $\alpha=[\alpha^{x},\alpha^{y}]$ represents the memory level, $\bar{v}=[\bar{v}^{x},\bar{v}^{y}]$ denotes the asymptotic mean, and $\bar{\kappa}=[\bar{\kappa}^{x},\bar{\kappa}^{y}]$ represents the asymptotic standard deviation of the velocity. Thus, given the velocity of $v_{k}(t)$, the position $\mathcal{P}_{k}(t)=[x_{k}(t),y_{k}(t)]$ at time slot $t$ of ED $u_k$ can be updated as $\mathcal{P}_{k}(t)=\mathcal{P}_{k}(t-1)+v_{k}(t)\Delta T.$
%\footnote{\textcolor{blue}{The reason for considering the single sub-channel in modelling is that we would like to specifically explore the ability of IRS that can mitigate the co-channel interference caused by NOMA transmission.}}. 

The IRS in the system has $M$ passive reflecting elements with the reflection coefficients matrix as $\boldsymbol{\varTheta}=diag\{\varTheta_1,\varTheta_2,...,\varTheta_M\}=diag\{\lambda_1e^{j\theta_1},\lambda_1e^{j\theta_2},...,\lambda_1e^{j\theta_M}\}$, where $\lambda_{m}\in[0,1]$ and $\theta_{m}\in [0,2\pi]$ are amplitude and phase shift of the $m^{th}$ reflection element on IRS\cite{yang2020intelligent}.
The equivalent baseband time-domain channel consists of three parts in the system: ED-IRS, ED-BS, and IRS-BS. To be specific, the channel gain between ED $u_k$ and the IRS is $\boldsymbol{g}_{k}\in \mathbb{C}^{1\times M}$, the channel gain between ED $u_k$ and the BS is $h^{d}_{k}$, and the channel gain between IRS and BS is $\boldsymbol{h}^{r}\in\mathbb{C}^{1\times M}$. Moreover, in practice, the IRS is installed near the EDs and may be located far away from BS. Hence, the ED-IRS channels would be LoS dominant that can be denoted as $\boldsymbol{g}_{k}=\sqrt{\beta_{k}^\text{(DI)}}\tilde{\boldsymbol{g}}_{k}$ with $\beta_{k}^\text{(DI)}$ as the large-scale path loss and $\tilde{\boldsymbol{g}}_{k}$ as the deterministic LoS channel between ED $u_k$ and the IRS. Differently, EDs may be far away from the BS with rich scatterers. Thus, we assume that the ED-BS channels are Rayleigh fading, which can be written as $h^{d}_{k}=\sqrt{\beta_{k}^{\text{(DB)}}}\tilde{h}^{d}_{k}$ with large-scale fading coefficient $\beta_{k}^{\text{(DB)}}$ and small-scale fading channel $\tilde{h}^{d}_{k}$. Furthermore, the IRS-BS channel can be modeled as Rician fading, which can be expressed as
\begin{align}
    \boldsymbol{h}^{r}=\sqrt{\beta^\text{(IB)}}\left(\sqrt{\frac{\delta}{1+\delta}}\tilde{\boldsymbol{h}}^{r\text{(LoS)}}+\sqrt{\frac{1}{1+\delta}}\tilde{\boldsymbol{h}}^{r\text{(NLoS)}}\right),
\end{align}
with $\beta^\text{(IB)}$ as the distance-dependent path loss factor between IRS and BS, $\delta$ as the Rician factor\footnote{When Rician factor $\delta=0$, the Rician fading becomes to Rayleigh fading. On the contrary, when $\delta\rightarrow{\infty}$, there is only LoS component.} which represents the ratio between LoS component $\tilde{\boldsymbol{h}}^{r,\text{(LoS)}}$ and non-LoS (NLoS) component $\tilde{\boldsymbol{h}}^{r,\text{(NLoS)}}$.

%SIC Decoding and achievable rate.
Conventionally, for NOMA systems, SIC decoding order is very crucial which is determined by the channel gains. In IRS-aided system, the IRS reflection coefficients can affect the combined channel gains. When the signal from ED $u_k$ needs to be decoded at the receiver BS, the achievable capacity
%\cite{liu2021energy} 
 in our proposed system for ED $u_k$, when prior decoding order were decoded, can be formulated as
\begin{align}
    R_{k}=B^{'}\text{log}_2\left(1+\frac{p_{k}\left|{h}_{k}\right|^2}{I_{k}+\sigma^2}\right),
    \label{achievablerate}
\end{align}
%where $\psi_{k}\in \{0,1\}$ indicates whether the $n^{th}$ channel is assigned to ED $u_k$, 
where $p_{k}$ is the power allocated to ED $u_{k}$, 
%use small case h. 
${h}_{k}= h_{k}^d+\boldsymbol{h}^{r}\boldsymbol{\varTheta}\boldsymbol{g}_{k}$ is the offloading channel which consists of both ED-BS and cascaded ED-IRS-BS paths, $I_{k}$ is the interference of the signals from EDs whose decoding order is latter than $k$, and the noise is assumed to be white Gaussian channel noise with variance as $\sigma^2$. Specifically,
$I_{k}=\sum_{\{q\in K| \eta(q)>\eta(k)\}}\left|{h}_{q}\right|^2 p_{q},$
%\textcolor{red}{add mapping function}
% eta???
% k tilde should be a set!!!--> how to reprent them separately.
%the channel gain information is known.--> known in advance..,--> CSI 
%\textcolor{red}{}
%with
% rou to be added in  the constraints.  
% equation (3)
%\begin{align}
%P_{n,\tilde{k}}^{\psi}=\sum_{q\rightarrow{}\tilde{k}}\psi_{n,q}p_{n,q},  
%\end{align}
which represents the overall interference of the rest EDs whose decoding order is latter than ED $u_k$  in this sub-channel (i.e., $\eta(q)$ and $\eta(k)$ are the decoding order mapping\footnote{In this paper, we deploy the dynamic decoding order mapping according to the channel gain, and it is assumed that the decoding order is well-known at the BS.} for ED $u_q$ and ED $u_k$, respectively. Thus, $\eta(q)>\eta(k)$ denotes that ED $u_q$ has the decoding order latter than ED $u_k$).
Therefore, the overall system communication throughput can be formulated as $R=\sum_{k=1}^{K}R_{k}$.
\subsection{Task Execution Model}
Since the main purpose of MEC is that each ED can execute its intensive tasks locally or offload to the edge server who has powerful computation ability. Hence, this subsection formulates the local computation model and edge computing model, respectively. 

\subsubsection{Local Computing}
When ${O}_{k}^{(l)}(t)=1$, ED $u_k$ executes task locally at time slot $t$, the computing energy consumption at each time slot $t$ can be formulated as
$E_{k}^{(l)}(t)=\pi^{(l)}f_{k}^{(l)}(t)^{3},$
where $\pi^{(l)}$ is the effective capacitance coefficient of processor chip\cite{8904347}. Moreover, the computing rate can be given as $R_{k}^{(l)}(t)={f_{k}^{(l)}(t)}/{c_{k}},$
where $f_{k}^{(l)}(t)$ indicates the CPU frequency on ED $u_k$, and $c_{k}$ is the CPU cycles required to process a single bit of data in cycles/bit for ED $u_k$. Therefore, the whole system energy consumption for local computing can be formulated as
$E^{(l)}(t)=\sum_{k\in \mathcal{K}^{(l)}(t)}E_{k}^{(l)}(t),$ with $\mathcal{K}^{(l)}(t)$ represents the EDs set who decide to execute the task locally. The system overall local computing rate can be denoted as
$R^{(l)}(t)=\sum_{k\in \mathcal{K}^{(l)}(t)}R_{k}^{(l)}(t).$
\subsubsection{Edge Computing}
% should be about user $k$
When ${O}_{k}^{(e)}(t)=1$, ED $u_k$ decides to offload the task for edge computing at time slot $t$. Compared to local EDs, edge server which is BS in this paper, has much powerful computation ability. The computing energy consumption at each time slot $t$ at BS can be formulated as $E^{(e)}(t)=\pi^{(e)}f^{(e)}(t)^{3},$ where $\pi^{(e)}$ is the effective capacitance coefficient of processor chip on BS. Moreover, the computing rate at BS can be given as $R^{(e)}(t)= f^{(e)}(t)/c_{0},$ where $f^{(e)}(t)$ indicates the CPU frequency allocated to process the offloaded tasks on BS, $c_{0}$ is the CPU cycles required to process a bit of data in cycles/bit on BS. 
%in equal bit. will it be improved??? to different edge computing user.
%c_{k}(t)
%The edge computing energy efficiency at AP can be given as
%\begin{align}
%    \varUpsilon^{(e)}(t)=R^{(e)}(t)/E^{(e)}(t).
%\end{align}
%Meanwhile, the edge computing energy efficiency of the $k^{th}$ ED can be denoted as $\varUpsilon^{(e)}_{k}(t)=\frac{f^{(e)}(t)/c_{k}(t)}{E^{(e)}_{t}}$. 
%The remaining CPU computation capacity ratio of the edge node AP can be denoted as $\mathcal{J}^{(edg)}(t)\in\{j_1^{(edg)},...,1\}$.
%Therefore, the computation latency of the task executed at AP is 
%\begin{align}
%L_{k}^{(E)}(t)=A_{k}(t) c_{k}/f.
%\end{align}
%Similar to local computing, the energy consumption by processing the offloaded task at AP can be formulated as
%\begin{align}
%    E_{k}^{(E)}(t)=A_{k}(t) c_{k} \pi^{(E)}
%\end{align}
%with $\pi^{(E)}$ as the computation capacity in terms of the CPU cycles per second at AP.

%Therefore, the cost function of the edge computing can be presented as
%\begin{align}
%C_{k}^{(\text{edg})}(t)=E_{k}^{(E)}(t)+\varpi_k  L_{k}^{(E)}(t). 
%\end{align}
%

\subsection{Task Offloading Communication Model}
Since different tasks generated by $u_k$ can be assigned to the same sub-channel with different transmit power, the achievable capacity in (\ref{achievablerate}) should be further presented as
\begin{align}
    R_{k}^{(o)}(t)=\min \left\{Q_{k}^{(l)}(t),B^{'}\text{log}_2\left(1+\frac{p_{k}(t)\left|{h}_{k}(t)\right|^2}{I_{k}(t)+\sigma^2}\right)\right\}
\end{align}
with time index $t$, $Q_{k}^{(l)}(t)$ represents the task queue size of ED $u_{k}$ at time slot $t$, and $p_{k}(t)$ represents the transmit power of ED $u_k$ at time slot $t$. The system overall arrival workloads at BS within a time slot can be written as
$R^{(o)}(t)=\sum_{k\in \mathcal{K}^{(e)}(t)}R_{k}^{(o)}(t),$
where $\mathcal{K}^{(e)}(t)$ denotes the subset of EDs who decide to offload the task at time slot $t$. The offloading transmission energy consumption can be given as
$E^{(o)}(t)=\sum_{k\in\mathcal{K}^{(e)}(t)}p_{k}(t)+M\varpi,$
where $M\varpi$ is the overall power consumption of IRS with $\varpi$ denoting the power consumption of each passive element on IRS \cite{9214497}.
%\textcolor{red}{should also consider the IRS power consumption}
%Therefore, the offloading transmission latency of the task $\varGamma_{k}(t)$ from ED $u_k$ to the AP with the help of IRS can be formulated as
%\begin{align}
%    L_{k}^{(O)}(t)=A_{k}(t)/R_{n,k}(t). 
%\end{align}
%Moreover, the energy consumption for computation tasks offloading can be written as
%\begin{align}
%    E_{k}^{(O)}(t)=p_{n,k}^{(O)}(t) L_{k}^{(O)}(t)
%\end{align}
%
%The offloading cost function can be written as
%\begin{align}
%C_{k}^{(\text{off})}(t)=E_{k}^{(O)}(t)+\varpi_k  L_{k}^{(O)}(t). 
%\end{align}
%The observed signal-to-interference-plus-noise ratio (SINR) for ED $u_{k}$ to the AP over IRS assisted channel $h_{k}(t)$ to offload task is formulated below as
%\begin{align}
%    \text{SINR}_{k}(t)=\frac{p_{k}(t)\left|{h}_{k}(t)\right|^2}{I_{k}(t)+\sigma^2}.
%\end{align}
During the offloading process, we assume that each ED adopts discrete transmit power control, with the transmit power values bounded by $\rho_{max}$. At time slot $t$, each end ED $u_k$ chooses its transmit power $p_{k}(t)$ from the range $[0,\rho_{max}]$ with different transmit power levels to determine whether to execute the task locally (i.e., $p_{k}(t)=0$) or to offload with selected power (i.e., $p_{k}(t)\in (0,...,\rho_{max}] $).

%discrete as vector $\mathcal{P}_{k}^{T}=\{\rho_{0},\rho_{1},...,\rho_{\mu}\}$ where $\rho_{0}=0$ denotes the ED process the task locally.

\subsection{Task Queue Model}
The task queue models at both ED $u_k$ and BS are illustrated in Fig. \ref{task}. The size of the tasks that can be processed at ED $u_k$ and BS during time slot $t$ can be respectively formulated as
\begin{align}
    \varLambda_{k}^{(l)}(t)=R_{k}^{(l)}(t)\Delta t=\frac{f^{(l)}_{k}(t)\Delta t}{c_{k}},
\end{align}
and
\begin{align}
    \varLambda^{(e)}(t)=R^{(e)}(t)\Delta t=\frac{f^{(e)}(t)\Delta t}{c_{0}}.
        %\varLambda^{(e)}(t)=R^{(e)}(t)\Delta t=\frac{f^{(e)}(t)\Delta t}{\sum_{k\in \mathcal{K}^{(es)}(t)}c_{k}},
\end{align}
Each ED $u_{k}$ has the local queue $Q_{k}^{(l)}(t)$ for unprocessed workloads. Similarly, the BS has a buffering offloaded queue $Q^{(e)}(t)$ to maintain the workloads received from EDs for edge computing. The input to $Q_{k}^{(l)}(t+1)$ for the next time slot $t+1$ consists of both generated task data and the remaining data, which can be updated as
\begin{equation}
\begin{split}
    Q_{k}^{(l)}(t+1) =\mathcal{G}_{k}(t)A_{k}(t)+[Q_{k}^{(l)}(t)-{O}_{k}^{(l)}(t)\varLambda_{k}^{(l)}(t)-{O}_{k}^{(e)}(t)R_{k}^{(o)}(t)\Delta t]^{+},
\end{split}
\end{equation}
%where $\eta_{n,k}\in\{0,1\}$ denotes whether the outage \footnote{The outage occurs when the receiving SINR of the user is below the required target SINR \cite{8649652}.} occurred or not (i.e., if outage happened, $\eta_{n,k}=0$ and the task offloading fails),
with  
%$\varLambda_{k}^{(l)}(t)$ is the amount of data processed by $u_{k}$ local CPU, 
$R_{k}^{(o)}(t)$ as the amount of data transmitted to BS during a time slot interval $\Delta T$ and $[\cdot]^+$ denotes $\max\{\cdot,0\}$. Moreover, the task queue at the BS can be updated as
\begin{equation}
    Q^{(e)}(t+1)=R^{(o)}(t)\Delta t+[Q^{(e)}(t)-\varLambda^{(e)}(t)]^{+},
\end{equation}
which is the combination of offloading workloads and the remaining data after edge computing.
% task queue-->[assum knwon to BS and EDs--> c_{k}]

\section{Problem Formulation}
\label{section3}
%overall description of the problem.
In this paper, we formulate the system overall energy efficiency maximization problem in the NOMA-based IRS-aided system by jointly optimizing the task offloading decisions $(\boldsymbol{O}^{(e)}$, $\boldsymbol{O}^{(l)})$, 
%SIC decode order $\boldsymbol{\varrho}=\{\varrho_{1},...,\varrho_{K}\}$, 
and transmit power of the EDs $\boldsymbol{p}=\{p_{1},...,p_{K}\}$, the CPU frequency on EDs $\boldsymbol{f}^{(l)}=\{f_1^{(l)},...,f_k^{(l)}\}$, the CPU frequency on BS $\boldsymbol{f}^{(e)}$, as well as the IRS reflection coefficients $\boldsymbol{\varTheta}$. We use $\mathcal{F}=\{\boldsymbol{O}^{(e)},\boldsymbol{O}^{(l)},\boldsymbol{p},\boldsymbol{f}^{(l)},\boldsymbol{f}^{(e)},\boldsymbol{\varTheta}\}$ to denote the whole solution set. 
\subsection{Problem Formulation}

%\textit{Definition 1:} 
The long-term overall energy efficiency $\Bar{\varUpsilon}$ is the sum of the system rate over the sum of the system energy consumption, which can be formulated as
\begin{equation}
\begin{split}
    \Bar{\varUpsilon}=
    \frac{\lim_{\tau\rightarrow\infty}\frac{1}{\tau}\sum_{t=0}^{\tau-1}\mathbb{E}\left[R(t)\right]}{\lim_{\tau\rightarrow\infty}\frac{1}{\tau}\sum_{t=0}^{\tau-1}\mathbb{E}\left[E(t)\right]}=\frac{\Bar{R}}{\Bar{E}},
    %\varUpsilon^{(l)}_{k}(t)+ \varUpsilon^{(e)}_{k}(t)+ \varUpsilon^{(o)}_{k}(t)-\xi_k(t).
   % E_k(t)=C_{k}^{(\text{loc})}(t)+C_{k}^{(\text{edg})}(t)+C_{k}^{(\text{off})}(t)+\xi(t)
    %\\&=E_{k}^{(L)}(t)+E_{k}^{(E)}(t)+E_{k}^{(O)}(t)+\varpi_k(  L_{k}^{(L)}(t)+L_{k}^{(E)}(t)+L_{k}^{(O)}(t))
\end{split}
\end{equation}
where $R(t)=R^{(l)}(t)+R^{(e)}(t)+R^{(o)}$ is the sum of the overall system rate at time slot $t$, and $E(t)=E^{(l)}(t)+E^{(e)}(t)+E^{(o)}(t)$ is the overall system energy consumption at time slot $t$.

%\textit{Remark 1:}
 %{Thus, the long-term system performance rather than the instantaneous performance can be defined as} 
%\begin{align}
%        \Bar{\varUpsilon}=\lim_{\mathcal{T}\rightarrow{\infty}}\frac{1}{\mathcal{T}}\sum_{t=0}^{T-1}\mathbb{E}\left[\varUpsilon(t)\right],
%\end{align}
%which is the time average expectation\cite{9463403} of the discrete-time system energy efficiency.
\begin{definition}
A discrete-time queue $Q(t)$ is mean rate stable\cite{neely2010stochastic} if
\begin{align}
    \lim_{t\rightarrow{\infty}}\frac{\mathbb{E}\left[Q(t)\right]}{t}=0.
\end{align}
\end{definition}
\begin{remark}
As shown in \textit{Definition 1}, task queue stability is guaranteed if the length of the queue is finite. According to the Little's Theorem, the system average delay is proportional to the average queue length based on the given traffic arrival rate. Therefore, queue stability is a crucial constraint to describe the system service delay.
\end{remark} 

Hence, the optimization problem can be formulated as
%\textcolor{red}{If we consider the partial offloading, then the constraint better be power related rather than computation frequency related then. Specially if user prefer to do offloading and local computation at the same time, the power consumption or the power ability is very important. As both local and offloading at the same time will definitely be more power-consumption.}
\begin{subequations}
\begin{align}
    (\textbf{P0}) \quad
    &\max_{\mathcal{F}}\quad\Bar{\varUpsilon}
    \label{a} \\
    \textbf{s.t.} \quad &O_{k}^{(e)}(t)\in \{0,1\},\quad \forall k\in \mathcal{K},
    \label{b}\\
    &O_{k}^{(l)}(t)\in \{0,1\},\quad \forall k\in \mathcal{K},
    \label{b2}\\
    &0 \leqslant p_{k}(t) \leqslant \rho_{max},\quad \forall k\in \mathcal{K}^{(e)}(t),
   \label{d}\\
    &0\leqslant f_{k}^{(l)}(t)\leqslant f_{k,max}^{(l)},\quad \forall k\in \mathcal{K}^{(l)}(t),
    \label{h}\\
    &0\leqslant f^{(e)}(t)\leqslant f_{max}^{(e)},
    \label{i}\\
    &\left|\varTheta_m(t) \right|\leqslant 1, \quad \forall m,
    \label{e}\\
    &\lim_{t\rightarrow{\infty}}\frac{\mathbb{E}\left[Q_{k}^{(l)}(t)\right]}{t}=0,\lim_{t\rightarrow{\infty}}\frac{\mathbb{E}\left[Q^{(e)}(t)\right]}{t}=0.\quad \forall k\in \mathcal{K}.\label{g}
\end{align}
\end{subequations}
%\textcolor{red}{Adjust the constraints sequences.}
Constraints (\ref{b},\ref{b2}) denote the offloading decision set and local computing decision set. 
%Constraint (\ref{c}) indicates the value range of weight factor $\varpi$ to balance the power consumption and execution latency. 
 %Constraint (\ref{f}) denotes that each channel can be assigned to $K_n$ users each time slot $t$. %Constraint (\ref{g}) restricts that each user can only be allocated to one channel each time slot $t$.  
 %Constraint (\ref{j}) represents the decoding order is the integer from 1 to the number of total number of users assigned to this channel. 
 Constraint (\ref{d}) denotes the transmit power of offloading. Constraints (\ref{h}) and (\ref{i}) mean that the maximum CPU frequency constraints of both EDs and BS. Constraint (\ref{e}) restricts the IRS reflection coefficient. Constraint (\ref{g}) denotes that the task queues are guaranteed to be stable.
%(\ref{h}) and (\ref{i}) constraint that each user must execute its task either locally or offload to edge node within one time slot.
It is noticed that (\ref{a}) is a long-term stochastic optimization problem in which multiple decision variables are adaptive to the dynamic system. However, due to the high stochastic and unpredictable of the dynamic system, it is impossible to acquire comprehensive knowledge for offline optimization. 
%Thus the original problem will be decoupled into slot-based optimization sub-problems.
% cost and constrain optimization problem.
Moreover, (\textbf{P0}) is a mixed integer nonlinear programming problem with binary constrains $\boldsymbol{O}^{(e)},\boldsymbol{O}^{(l)}$, continuous variables $\boldsymbol{p},\boldsymbol{f}^{(l)},\boldsymbol{f}^{(e)}$, and $\boldsymbol{\varTheta}$. Therefore, it is even challenging and inefficient to find the optimal solution through conventional optimization techniques which decouple the problem into sub-problems and solve them separately. %To overcome these challenges, we explore the online DRL-based techniques which can efficiently make decisions in real-time through constantly interactions with the system.
\subsection{Problem Reformulation}
It is observed that the energy efficiency optimization problem in (\textbf{P0}) is a stochastic nonlinear programming problem, which is also a NP-hard problem with nonlinear fractional property. For notation simplicity, $\mathcal{F}^{*}=\{\boldsymbol{O}^{(e)*},\boldsymbol{O}^{(l)*},\boldsymbol{p}^{*},\boldsymbol{f}^{(l)*},\boldsymbol{f}^{(e)*},\boldsymbol{\varTheta}^{*}\}$ 
%=\{\mathcal{F}^{'*},\boldsymbol{\varTheta}^{*}\}$ 
denotes the feasibility solutions of (\textbf{P0}) with the optimal solution sequence $\{\mathcal{F}^{*}(t)_{t\in{0,1,2,...}}\}$ that can maximize $\Bar{\varUpsilon}$ from a long-term perspective under all constraints, which can be written as
\begin{align}
    \Bar{\varUpsilon}^{opt}=\max_{\mathcal{F}^{*}}\frac{\Bar{R}(\mathcal{F}^{*})}{\Bar{E}(\mathcal{F}^{*})}.
\end{align}
\begin{theorem} The optimal energy efficiency $\Bar{\varUpsilon}^{opt}$ can be achieved if and only if
\begin{equation}
\begin{split}
    \max_{\mathcal{F}^{*}}\quad \Bar{R}(\mathcal{F})-\Bar{\varUpsilon}^{opt}\Bar{E}(\mathcal{F})= \Bar{R}(\mathcal{F}^{*})-\Bar{\varUpsilon}^{opt}\Bar{E}(\mathcal{F}^{*})=0.
    \label{theory}
\end{split}
\end{equation}
\begin{proof}
This theorem can be proved from two aspects: the necessity and the sufficiency. Firstly, based on the necessity, for any feasible solutions $\mathcal{F}$, we have 
\begin{align}
    \Bar{\varUpsilon}^{opt}=\frac{\Bar{R}(\mathcal{F}^{*})}{\Bar{E}(\mathcal{F}^{*})}\geqslant {\frac{\Bar{R}(\mathcal{F})}{\Bar{E}(\mathcal{F})}}.
\end{align}
Then, the formula can be rearranged as
\begin{align}
    \Bar{R}(\mathcal{F})-\Bar{\varUpsilon}^{opt}\Bar{E}(\mathcal{F})\leqslant  0, \quad
    \Bar{R}(\mathcal{F}^{*})-\Bar{\varUpsilon}^{opt}\Bar{E}(\mathcal{F}^{*})=0,
\end{align}
which completes the necessity proof.
Secondly, based on sufficiency proof, we firstly assume that $\mathcal{F}^{'}$ is the optimal solution of (\ref{theory}). Thus, for any feasible solution, we have \begin{align}
    \Bar{R}(\mathcal{F})-\Bar{\varUpsilon}^{opt}\Bar{E}(\mathcal{F})\leqslant\Bar{R}(\mathcal{F}^{'})-\Bar{\varUpsilon}^{opt}\Bar{E}(\mathcal{F}^{'})=0.
\end{align}
Then, we can obtain
\begin{align}
    \frac{\Bar{R}(\mathcal{F})}{\Bar{E}(\mathcal{F})}\leqslant\Bar{\varUpsilon}^{opt}, \quad \frac{\Bar{R}(\mathcal{F}^{'})}{\Bar{E}(\mathcal{F}^{'})}=\Bar{\varUpsilon}^{opt}.
\end{align}
Therefore, $\mathcal{F}^{'}$ is the optimal solution of (\ref{a}), which completes the sufficient proof.
\end{proof}
\end{theorem}

Thus, with the support of \textbf{Theorem 1}, we further transform the energy efficiency maximization problem as
    \begin{align}
    (\textbf{P1})\quad&\max_{\mathcal{F}}\quad \Bar{R}-\Bar{\varUpsilon}^{opt}\Bar{E}\\
    \textbf{s.t.}\quad &(\ref{b}),(\ref{d}),(\ref{h}),(\ref{i}),(\ref{e}),(\ref{g}).    \nonumber
    \end{align}
However, the optimal value of $\Bar{\varUpsilon}^{opt}$ cannot be known in advance. To overcome such difficulty, a variable $\varUpsilon(t)$ is defined as
\begin{align}
    \varUpsilon(t)=\frac{\sum_{\tau=0}^{t-1}R(\tau)}{\sum_{\tau=0}^{t-1}E(\tau)},
    \label{varupsilont}
\end{align}
with $t\in\{1,2,...,T\}$ and $\varUpsilon(0)=0$. It can be observed that in (\ref{varupsilont}), $ \varUpsilon(t)$ can be treated as the parameter which is mainly determined by the past task offloading and resource allocation decisions. Then, the optimization problem (\textbf{P1}) can be reformulated as
\begin{align}
    (\textbf{P2})\quad &\max\quad \Bar{R}-\varUpsilon(t)\Bar{E}\\
    \textbf{s.t.}\quad &(\ref{b}),(\ref{d}),(\ref{h}),(\ref{i}),(\ref{e}),(\ref{g}).\nonumber
\end{align}
Similar to the transformation in \cite{6893054} on handling the stochastic optimization problems in renewal systems, it has been proved that the reformulated problem (\textbf{P2}) can effectively solve the original problem.
%Conventionally, deep Q-network (DQN) is an advanced RL algorithms which can handle high-dimensional discrete state spaces. Although the problem with continuous state and action spaces in our system can be transferred to discrete format, the dropping precision and increasing complexity makes it infeasible to apply DQN on solving our problem directly. Deep deterministic policy gradient (DDPG) has been proposed which can learn competitive policies with continuous action and state spaces \cite{lillicrap2015continuous}. In traditional DQN, a DNN is learned to represent the-Q function on helping determine the optimal policy. Differently, in DDPG, there are two different DNNs namely critic network $\mathcal{Q}(\boldsymbol{s},\boldsymbol{a}|\boldsymbol{\varPhi}_{\mathcal{Q}})$ which approximates the Q-function, and actor network $\chi(\boldsymbol{s}|\boldsymbol{\varPhi}_{\chi})$ which approximates the policy function $\chi$, respectively. To be noticed, $\boldsymbol{\varPhi}_{\mathcal{Q}}$ and $\boldsymbol{\varPhi}_{\chi}$ denote the weights of the critic and actor network DNNs.
%\subsection{Centralized Decision-Making: DDPG}
%Deep policy gradient.
%\subsection{Decentralized Decision-Making: multi-agent DDPG}

\subsection{Lyapunov Optimization}
As there is the task queue stability constraint (\ref{g}) in the optimization problem (\textbf{P2}), Lyapunov method can be adapted to solve the stochastic optimization problem and investigate the energy efficiency and task queue length trade-off. Therefore, with the system queue length $\boldsymbol{Q}(t)=\left[Q_1^{(l)}(t),...,Q_k^{(l)}(t),...,Q_K^{(l)}(t),Q^{(e)}{t}\right]=\left[Q_{1}(t),...,Q_{K+1}(t)\right]$ at time slot $t$, the Lyapunov-function and the Lyapunov drift can be written as
\begin{align}
    L\left(Q(t)\right)=\frac{1}{2}Q^{(e)}{(t)}^2+\frac{1}{2}\sum_{k\in \mathcal{K}}Q_k^{(l)}(t)^2,
\end{align}
and
\begin{align}
    \Delta L\left(Q(t)\right)=L\left(Q(t+1)\right)-L\left(Q(t)\right).
\end{align}
Furthermore, the Lyapunov drift-plus-penalty function at each time slot $t$ can be formulated as
\begin{equation}
\begin{split}
    \Delta L\left(\boldsymbol{Q}(t)\right)-V\left(\Bar{R}-\varUpsilon(t)\Bar{E}\right)=\underbrace{L\left(\boldsymbol{Q}(t+1)\right)-L\left(\boldsymbol{Q}(t)\right)}_{\text{drift}}+V\underbrace{\left(\varUpsilon(t)\Bar{E}-\Bar{R}\right)}_{\text{penalty}},
    \label{driftpluspenalty}
    \end{split}
\end{equation}
where $V$ is the positive weight control parameter to tune the trade-off between the drift and the penalty. To ensure the feasibility of the Lyapunov drift-plus-penalty optimization with finite queue capacity, the enforcement constraint has been proposed in \cite{9152999}. Specifically, this paper indicates that the queue stability capacity constraint can be realized by carefully adjust the factor $V$ value. In Lyapunov theorem, to minimize the function in eq. (\ref{driftpluspenalty}), the strategy can be refined by deriving an upper bound which can be provided as
\begin{equation}
\begin{split}
     &\Delta L\left(\boldsymbol{Q}(t)\right)-V\left(\Bar{R}-\varUpsilon(t)\Bar{E}\right)\\&\leqslant b+\mathbb{E}\left\{\sum_{j=1}^{K+1}Q_{j}(t)\left(Q_{j}^{(in)}(t)-Q_{j}^{(out)}(t)\right)|\boldsymbol{Q}(t)\right\}+V\mathbb{E}\left\{\varUpsilon(t)\Bar{E}-\Bar{R}|\boldsymbol{Q}(t\right\},
\end{split}
\end{equation}
where $Q_{j}^{(in)}(t)$ is the arrived data size at queue $Q_{j}(t)$ at time slot $t$, $Q_{j}^{(out)}(t)$ is the offloaded/processed data size at the $Q_{j}(t)$ at time slot $t$, and $b$ is the positive constant that upper bounds the expression $\mathbb{E}\left\{\sum_{j=1}^{K+1}\frac{1}{2}\left(Q_{j}^{(in)}(t)^2+Q_{j}^{(out)}(t)^2\right)|\boldsymbol{Q}(t)\right\}$. The proof can be found in Appendix A. Therefore, the optimization problem in (\textbf{P2}) can be achieved by minimizing 
\begin{equation}
    \begin{split}
    \mathcal{L}(t)=\sum_{j=1}^{K+1}Q_{j}(t)\left(Q_{j}^{(in)}(t)-Q_{j}^{(out)}(t)\right)+V\left(\varUpsilon(t)\Bar{E}-\Bar{R}\right).
    \label{upperbound}
    \end{split}
\end{equation}

%In another words, the weighted difference in the queue length plus the weighted penalty should be minimized.
%\begin{equation}
%\begin{split}
%    &(\textbf{P3})\quad \min\quad \sum_{j=1}^{K+1}Q_{j}(t)\left(Q_{j}^{(in)}(t)+Q_{j}^{(out)}(t)\right)+\\&V\left(\varUpsilon(t)\Bar{E}-\Bar{R}\right)
%    \\&\textbf{s.t.}\quad (\ref{b}),(\ref{d}),(\ref{h}),(\ref{i}),(\ref{e}).
%\end{split}
%\end{equation}
\section{Proposed Algorithm}
\label{section4}
Deep Q-network (DQN) is an advanced RL algorithms which can handle high-dimensional discrete state spaces. Although the problem with continuous action spaces in our system can be transferred to discrete format through quantization, the dropping precision and significantly increased complexity makes it infeasible to exploit DQN to solve our problem directly. Actor-critic RL structure has been proposed to solve the problem with continuous action spaces. This is because the actor network outputs an action through the actor deep neural network (DNN) at each time step and the critic network evaluates the reward or the Q-value of a given input state. As the critic network learns which states are better or worse, the actor uses this information to teach the agent to seek out good states and avoid bad states. There are three popular algorithms under this actor-critic structure: Proximal Policy Optimization (PPO)\cite{schulman2017proximal}, advantage actor critic (A2C) algorithm\cite{mnih2016asynchronous} and deep deterministic policy gradient (DDPG)\cite{lillicrap2015continuous}. Among three actor-critic based algorithms, PPO is more sample expensive which is because the set of trajectories should be obtained by running the current policy over certain time slots. However, after the policy being updated, the old trajectories generated based on previous policy are no longer applicable for training the network. Thus, in this paper, considering the dynamic IRS-assisted NOMA-aided MEC work, it's more practical to consider the sample efficient algorithms. The main difference between A2C and DDPG is the way they apply critic network. In A2C, the critic is treated as a baseline from empirical trajectories, so whether to use only critic is just an option to improve the robustness of the training as the random selection prevents the back-propagation for training. However, in DDPG, the policy is deterministic and the gradient can be acquired from the Q-value, where Q-value is obtained from the critic network and the actions generated by the actor network. More importantly, DDPG is off-policy which means that many history trajectories can be utilized for training.

In order to adapt RL-based techniques to solve (\textbf{P2}), we reformulate the problem into the RL framework which consists of a few key elements, such as agent \footnote{It's worth noticing that, for centralized scheme, the agent is the BS. For distributed scheme, there are heterogeneous multiple agents which are EDs and BS.}, environment, states $\mathcal{S}$, actions $\mathcal{A}$, and reward $\mathcal{R}$. Typically, the environment is defined as the Markov Decision Process (MDP). Although dynamic programming is a classic solution on solving MDP problems, when encounters large-scale MDP problems, it is infeasible to utilize dynamic programming as it requires the knowledge of the mathematical model (i.e., Markovian transition probabilities). Generally, in each time slot $t$, the RL agent can continually interact with the environment and observe the state $\boldsymbol{s}(t)\in \mathcal{S}$, then it takes an action $\boldsymbol{a}(t)\in\mathcal{A}$. The agent's behavior follows the rule of a policy $\chi$ (i.e., $\chi: \mathcal{S}\rightarrow \mathcal{A}$). In return, the environment calculates the reward $\boldsymbol{r}(t)=\mathcal{R}(\boldsymbol{s}(t),\boldsymbol{a}(t))$ and changes the state from $\boldsymbol{s}(t)$ to $\boldsymbol{s}(t+1)$. 
In other words, the transition probability  $\mathbb{P}(\boldsymbol{s}(t+1),\boldsymbol{r}(t)| \boldsymbol{s}(t),\boldsymbol{a}(t))$ of the MDP which is defined as the transition from state $\boldsymbol{s}(t)$ to $\boldsymbol{s}(t+1)$ with reward $\boldsymbol{r}(t)$ when the action $\boldsymbol{a}(t)$ is taken according to the policy.

In this section, we firstly propose the centralized framework called LMIDDPG which can solve the problem (\textbf{P2}) with the mixed integer variables as the action space. At the same time, the negative of the upper bound Lyapunov drift-plus-penalty function has been treated as the reward function. In order to further save the communication cost at both training and execution stages, we then propose the distributed framework called HMA-LMIDDPG with heterogeneous multi-agent.
\subsection{Centralized RL Framework: Lyapunov-Function-Based Mixed Integer Deep Deterministic Policy Gradient}
\begin{figure}[t]
  \centering
  \includegraphics[width=6.5in]{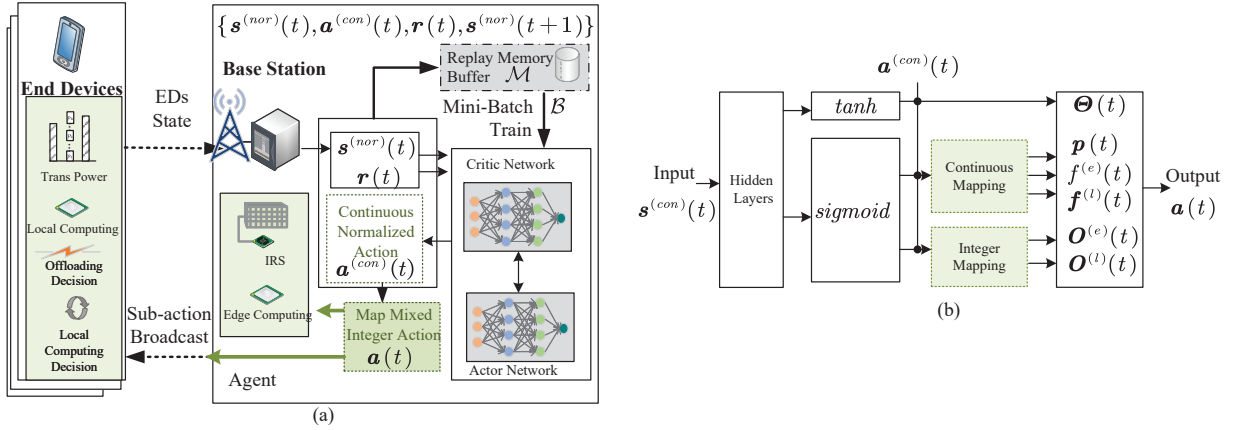}\\
   \caption{(a) The training stage of the centralized RL framework: LMIDDPG, (b) the actor network structure with mixed integer mapping at centralized agent BS.}
  \label{fig:ddpg}
\end{figure}
In fundamental DDPG, the two different DNNs namely critic network $\mathcal{Q}(\boldsymbol{s},\boldsymbol{a}|\boldsymbol{\varPhi}_{\mathcal{Q}})$ which approximates the Q-function, and actor network $\chi(\boldsymbol{s}|\boldsymbol{\varPhi}_{\chi})$ which approximates the policy function $\chi$, respectively. To be noticed, $\boldsymbol{\varPhi}_{\mathcal{Q}}$ and $\boldsymbol{\varPhi}_{\chi}$ denote the weights of the critic and actor network DNNs. 
The process of centralized training and execution framework is shown in Fig. \ref{fig:ddpg}(a). Under this framework, the BS is the agent that can observe the environment by receiving the EDs states. With this interaction, BS will obtain the experience buffer memory at each time slot $t$ to train the critic network $\mathcal{Q}(\boldsymbol{s},\boldsymbol{a}|\boldsymbol{\varPhi}_{\mathcal{Q}})$ and the actor network $\chi(\boldsymbol{s}|\boldsymbol{\varPhi}_{\chi})$. During the training and execution stages, EDs receive the sub-action instructions which are determined by the actor network from the BS. Before giving the details of the training process of the proposed LMIDDPG algorithm, the design of the centralized agent and mixed integer need to be explained:
\subsubsection{Centralized Agent}
 In the centralized RL framework, the BS acts as the agent. The details definition of the state space, action space, and the reward are defined as follows:
\begin{itemize}
    \item \textbf{State space}: The state $\boldsymbol{s}(t)$ at each time slot $t$ can be defined as $\boldsymbol{s}(t)=\{R^{(o)}(t),Q(t),Q(t-1),\mathcal{P}(t)\}$ with the information acquired from the EDs and BS itself. 
    \item \textbf{Action space}: The action vector of the whole system can be formulated as $\boldsymbol{a}(t)=\{\mathcal{F}(t)\}$ with mixed integer property.
    \item \textbf{Reward}: The reward function of the whole system at time slot $t$ can be defined as $\boldsymbol{r}(t)=\mathcal{R}(\boldsymbol{s}(t),\boldsymbol{a}(t))=-\mathcal{L}(t),$ which is the negative value of the function derivated from the upper-bound Lyapunov drift-plus-penalty function in eq. (\ref{upperbound}).
\end{itemize}
\subsubsection{Mixed Integer Mapping}
DDPG algorithm is an extension version of the deterministic policy gradient which approximates the actor and critic functions based on deep neural networks (DNNs) that can learn policies in high-dimensional, continuous action spaces \cite{lillicrap2015continuous}. However, in our problem, the action space is a mixed integer space with integer sub-actions: offloading decision ${O}^{(e)}_{k}(t)\in \{0,1\}$, local computing decision ${O}_{k}^{(l)}(t)\in \{0,1\}$, and continuous sub-actions: transmit power ${p}_{k}(t)\in [0,\rho_{k,max}]$, local computing frequency ${f}_{k}^{(l)}(t)\in [0,f_{k,max}^{(l)}]$, edge computing frequency ${f}^{(e)}(t)\in [0,f_{max}^{(e)}]$, and IRS phase shifts ${\varTheta}(t)\in[-1,1]$. In the proposed framework, the actor network outputs the normalized continuous action space $\boldsymbol{a}^{(con)}(t)$
$\boldsymbol{a}^{(con)}(t)=\{\boldsymbol{O}^{(e)'},\boldsymbol{O}^{(l)'},\boldsymbol{p}^{'},\boldsymbol{f}^{(l)'},\boldsymbol{f}^{(e)'},\boldsymbol{\varTheta}\}$ that cannot be directly executed in the environment. Thus, a mapping procedure is designed to recover the true mixed integer action space $\boldsymbol{a}(t)$. The structure of the actor network and mapping details can be found in Fig. \ref{fig:ddpg}(b). After different activation function (i.e., $\boldsymbol{\varTheta}$ is activated through function $tanh$, the rest sub-actions are activated through $sigmoid$), the sub-actions $\boldsymbol{p}_{k}^{'},\boldsymbol{f}_{k}^{(l)'},\boldsymbol{f}^{(e)'}$ in the normalized continuous action space can be further mapped to the true value by multiplying their individual maximum value ${\rho}_{k,max},{f}_{k,max}^{(l)},\boldsymbol{f}^{(e)}_{max}$. Additionally, both offloading decision ${O}^{(e)}_{k}(t)$ and local computing decision ${O}^{(l)}_{k}(t)$ will pass through the $1/2$ threshold and mapped to the binary integer space $\{0,1\}$. 

\begin{algorithm}[!t]
\small
\label{algo1}
 \caption{Lyapunov-Function-Based Mixed Integer Deep Deterministic Policy Gradient (LMIDDPG)}
 \hspace*{\algorithmicindent}
\textbf{Initialization:} Actor (policy) network $\chi(\boldsymbol{s}|\boldsymbol{\varPhi}_{\chi})$ with parameters $\boldsymbol{\varPhi}_{\chi}$, critic (Q-function) network $\mathcal{Q}(\boldsymbol{s},\boldsymbol{a}|\boldsymbol{\varPhi}_{\mathcal{Q}})$ with parameters $\boldsymbol{\varPhi}_{\mathcal{Q}}$, target actor network $\chi^{'}$ with parameters $\boldsymbol{\varPhi}_{\chi^{'}}\leftarrow{\boldsymbol{\varPhi}_{\chi}}$, target critic network $\mathcal{Q}^{'}$ with parameters $\boldsymbol{\varPhi}_{\mathcal{Q}^{'}}\leftarrow{\boldsymbol{\varPhi}_{\mathcal{Q}}}$, empty replay buffer memory $\mathcal{M}$.  
 \begin{algorithmic}[1]
 \FOR {episode $=1$ to $\mathcal{E}$}
\STATE Initialize environment, i.e., initial observation state $\boldsymbol{s}(1)$.
 \FOR{time $t=1$ to $T$}
   \STATEx{{\textbf{\% Mixed integer action space mapping:}}}
 \STATE Observe current normalized system state $\boldsymbol{s}^{(nor)}(t)$ and choose normalized continuous action $\boldsymbol{a}^{(con)}(t)$ by the actor network. 
 \STATE Map the normalized continuous action $\boldsymbol{a}^{(con)}(t)$ into true mixed integer action space $\boldsymbol{a}(t)$. 
 %  \STATEx{{\textbf{//Enforce the finite queue size constraint:}}}
 %\IF {$Q_{j\rightarrow{[1,K+1]}}(t)>Q_{j\rightarrow{[1,K+1]}}^{th}$}
Execute $\boldsymbol{a}(t)$ in the environment.
\STATEx {\textbf{\% Lyapunov queue stability award training: }}
 \STATE Observe the Lyapunov-function-based reward $\boldsymbol{r}(t)$ and new state $\boldsymbol{s}(t)$. 
 \STATE {Normalize new state as $\boldsymbol{s}^{(nor)}(t+1)$. Store $\{\boldsymbol{s}^{(nor)}(t),\boldsymbol{a}^{(con)}(t),\boldsymbol{r}(t),\boldsymbol{s}^{(nor)}(t+1)\}$ in the replay buffer memory $\mathcal{M}$.}
 \STATE Sample a random mini-batch $\mathcal{B}$ transitions $\{\boldsymbol{s}^{(nor)}(t),\boldsymbol{a}^{(con)}(t),\boldsymbol{r}(t),\boldsymbol{s}^{(nor)}(t+1)\}$ from $\mathcal{M}$.
  \STATE Compute $y=\boldsymbol{r}(t)+\gamma\mathcal{Q}^{'}\left(\boldsymbol{s}^{(nor)}\left(t+1\right),\chi^{'}\left(\boldsymbol{s}^{(nor)}(t+1)|\boldsymbol{\varPhi}_{\chi^{'}}\right)|\boldsymbol{\varPhi}_{\mathcal{Q}^{'}}\right)$
  \STATE Update critic by minimizing the loss: $\mathcal{L}^{'}=\frac{1}{\mathcal{B}} \sum\left(y-\mathcal{Q}(\boldsymbol{s}^{(nor)}(t),\boldsymbol{a}^{(con)}(t)|\boldsymbol{\varPhi}_{\mathcal{Q}})\right)^2$ 
  %$\mathcal{L}=\frac{1}{\mathcal{B}} \sum_{\boldsymbol{s}(t),\boldsymbol{a}(t),\boldsymbol{r}(t),\boldsymbol{s}(t+1) \in \mathcal{B}}\left(y-\mathcal{Q}(\boldsymbol{s}(t),\boldsymbol{a}(t)|\boldsymbol{\varPhi}_{\mathcal{Q}})\right)^2$ 
  \STATE Update actor policy using the sampled policy gradient:
  $\nabla_{\boldsymbol{\varPhi}_{\chi}}\frac{1}{\mathcal{B}}\sum_{\boldsymbol{s} \in \mathcal{B}}  \mathcal{Q}(\boldsymbol{s}^{(nor)}(t),\chi(\boldsymbol{s}^{(nor)}(t)|\boldsymbol{\varPhi}_{\chi})|\boldsymbol{\varPhi}_{\mathcal{Q}})$
\STATE Update target networks with $\boldsymbol{\varPhi}_{\chi^{'}}\leftarrow{\varrho\boldsymbol{\varPhi}_{\chi}+(1-\varrho)\boldsymbol{\varPhi}_{\chi^{'}}}$ and $\boldsymbol{\varPhi}_{\mathcal{Q}^{'}}\leftarrow{\varrho\boldsymbol{\varPhi}_{\mathcal{Q}}+(1-\varrho)\boldsymbol{\varPhi}_{\mathcal{Q}^{'}}}$
\ENDFOR
\ENDFOR
\end{algorithmic}
\end{algorithm}
%training procedure. 
As LMIDDPG is designed based on the centralized RL framework, there is only one critic network and one actor network, where the actor network can output the complete action space with multiple tasks. The training process pseudocode of the proposed algorithm is given in \textbf{Algorithm 1}. For each time slot $t$, the normalized state $\boldsymbol{s}^{(nor)}(t)$, normalized continuous action $\boldsymbol{a}^{(con)}(t)$, reward $\boldsymbol{r}(t)$, and the normalized next state $\boldsymbol{s}^{(nor)}(t+1)$ will be stored in the replay buffer memory $\mathcal{M}$ at BS. Then, during the training process, a mini-batch $\mathcal{B}$ will be sampled from the buffer memory $\mathcal{M}$ to compute the target value
\begin{equation}
    y=\boldsymbol{r}(t)+\gamma\mathcal{Q}^{'}\left(\boldsymbol{s}^{(nor)}\left(t+1\right),\chi^{'}\left(\boldsymbol{s}^{(nor)}(t+1)|\boldsymbol{\varPhi}_{\chi^{'}}\right)|\boldsymbol{\varPhi}_{\mathcal{Q}^{'}}\right),
\end{equation}
where $\gamma\in [0,1]$ is the discounting factor, $\mathcal{Q}^{'}$ indicates the Q-value calculated based on the target critic network, $\chi^{'}\left(\boldsymbol{s}^{(nor)}(t+1)|\boldsymbol{\varPhi}_{\chi^{'}}\right)$ indicates the normalized continuous action instructed by the target actor network ${\chi^{'}}$ given the next normalized state $\boldsymbol{s}^{(nor)}(t+1)$. Then, the critic network parameters can be updated based on a learning rate $\varrho_{\mathcal{Q}}$ by minimizing the loss function 
\begin{equation}
\mathcal{L}^{'}=\frac{1}{\mathcal{B}} \sum\left(y-\mathcal{Q}(\boldsymbol{s}^{(nor)}(t),\boldsymbol{a}^{(con)}(t)|\boldsymbol{\varPhi}_{\mathcal{Q}})\right)^2.
\end{equation}
In DDPG, the actor network is approximated by a DNN and the policy gradient of the actor can be updated by applying the chain rule to the reward with respect to the actor network parameters $\boldsymbol{\varPhi}_{\mathcal{X}}$. The sampled policy gradient can be derived as
\begin{equation}
    \nabla_{\boldsymbol{\varPhi}_{\chi}}\frac{1}{\mathcal{B}}\sum_{\boldsymbol{s} \in \mathcal{B}}  \mathcal{Q}\left(\boldsymbol{s}^{(nor)}(t),\chi\left(\boldsymbol{s}^{(nor)}(t)|\boldsymbol{\varPhi}_{\chi}\right)|\boldsymbol{\varPhi}_{\mathcal{Q}}\right).
\end{equation}
Then, the actor network parameter can be updated based on the learning rate $\varrho_{\chi}$. To improve the stability of the learning process, the weights of both target actor network and target policy network are then updated by slowly tracking the learned parameters: $\boldsymbol{\varPhi}_{\chi^{'}_{i}}\leftarrow{\varsigma\boldsymbol{\varPhi}_{\chi_{i}}+(1-\varsigma)\boldsymbol{\varPhi}_{\chi^{'}_{i}}}$ and $\boldsymbol{\varPhi}_{\mathcal{Q}^{'}_{i}}\leftarrow{\varsigma\boldsymbol{\varPhi}_{\mathcal{Q}_{i}}+(1-\varsigma)\boldsymbol{\varPhi}_{\mathcal{Q}^{'}_{i}}}$ with $\varsigma$ as the update rate.

\subsection{Distributed RL Framework: Heterogeneous Multi-agent Lyapunov Mixed Integer Deep Deterministic Policy Gradient}
\begin{figure*}[t]
  \centering
  \includegraphics[width=6.6in]{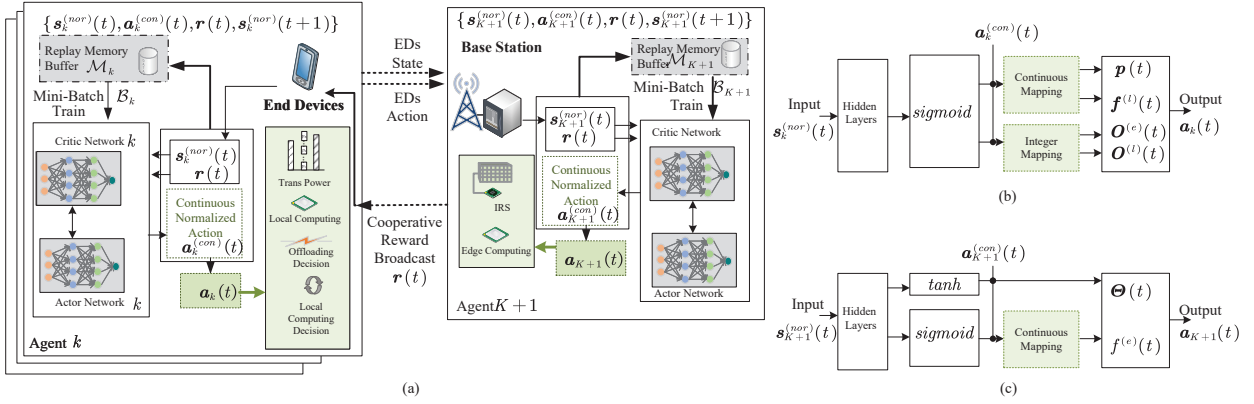}\\
   \caption{(a) The training stage of the distributed RL framework: HMA-LMIDDPG, (b) the actor network structure with mixed integer mapping at agent ED $u_{k}$, (c) the actor network at agent BS.}
  \label{fig:maddpg}
\end{figure*}
\begin{algorithm}[!t]
\small
\label{algo2}
 \caption{Heterogeneous Multi-agent Lyapunov-Function-Based Mixed Integer Deep Deterministic Policy Gradient (HMA-LMIDDPG)}
 \hspace*{\algorithmicindent}
\textbf{Initialization:} Agent ${i}$ ($i\in\{1,...,K+1\}$. When $i=K+1$, the agent is BS. Otherwise, the agent is ED $u_{k}$.) Actor network $\chi_{i}(\boldsymbol{s}_{i}|\boldsymbol{\varPhi}_{\chi_{i}})$ with parameters $\boldsymbol{\varPhi}_{\chi_{i}}$, Agent ${i}$ critic network $\mathcal{Q}_{i}(\boldsymbol{s}_{i},\boldsymbol{a}_{i}|\boldsymbol{\varPhi}_{\mathcal{Q}_{i}})$ with parameters $\boldsymbol{\varPhi}_{\mathcal{Q}_{i}}$, target Agent ${i}$ actor network $\chi^{'}_{i}$ with parameters $\boldsymbol{\varPhi}_{\chi^{'}_{i}}\leftarrow{\boldsymbol{\varPhi}_{\chi_{i}}}$, target Agent ${i}$ critic network $\mathcal{Q}_{i}^{'}$ with parameters $\boldsymbol{\varPhi}_{\mathcal{Q}^{'}_{i}}\leftarrow{\boldsymbol{\varPhi}_{\mathcal{Q}_{i}}}$, empty replay buffer memory $\mathcal{M}_{i}$ at each Agent ${i}$.  
 \begin{algorithmic}[1]
%\STATEx{{\textbf{\%Generate the complete agent list:}}}
% \FOR{Agent $i=0$ to $K$}
%\STATE {agentlist.append(Agent ${i}$)}
%\ENDFOR
\FOR {Episode $=1$ to $\mathcal{E}$}
\STATEx{{\textbf{\%Heterogeneous multi-agent:}}}
\STATE {Each Agent $i$ has its own initial observation state $\boldsymbol{s}_{i}(1)$. Specially, compared to homogeneous EDs as the Agent $1$ to Agent $K$, BS as the Agent $K+1$ has different dimension of state space and action space.} 
%\STATE {Normalize the overall statelist $\boldsymbol{s}_{all}(1)=\{\boldsymbol{s}_{0}(1),\boldsymbol{s}_{1\rightarrow{K}}(1)\}$.}
 \FOR{Time $t=1$ to $T$}
 \FOR{Agent $i=1$ to $K+1$}
 \STATE Observe current normalized system state $\boldsymbol{s}_{i}^{(nor)}(t)$ and choose normalized continuous action $\boldsymbol{a}_{i}^{(con)}(t)$ by the actor network $\chi_{i}(\boldsymbol{s}_{i}|\boldsymbol{\varPhi}_{\chi_{i}})$.  
\STATE{Map the normalized continuous action $\boldsymbol{a}_{i}^{(con)}(t)$ into true mixed integer action space $\boldsymbol{a}_{i}(t)$. Execute $\boldsymbol{a}_{i}(t)$ in the environment.}
 \ENDFOR
\STATEx{{\textbf{\%Cooperative Lyapunov queue stability award training:}}}
 \STATE {Observe the cooperative Lyapunov-Function-Based reward $\boldsymbol{r}(t)$ based on the statelist $\boldsymbol{s}_{all}(t)=\{\boldsymbol{s}_{1\rightarrow{K}}(t),\boldsymbol{s}_{K+1}(t)\}$.
 and actionlist $\boldsymbol{a}_{all}(t)=\{\boldsymbol{a}_{1\rightarrow{K}}(t),\boldsymbol{a}_{K+1}(t)\}$. 
 Broadcast the cooperative reward to all EDs.}
 \FOR{Agent $i=1$ to $K+1$}
 \STATE {Observe the new state $\boldsymbol{s}_{i}(t+1)$, and normalize the new state as $\boldsymbol{s}_{i}^{(nor)}(t+1)$.}
 \STATE {Store $\{\boldsymbol{s}_{i}^{(nor)}(t),\boldsymbol{a}_{i}^{(con)}(t),\boldsymbol{r}(t),\boldsymbol{s}_{i}^{(nor)}(t+1)\}$ in the replay buffer memory $\mathcal{M}_{i}$.}
 \STATE {Sample a random mini-batch $\mathcal{B}_{i}$ transitions $\{\boldsymbol{s}_{i}^{(nor)}(t),\boldsymbol{a}_{i}^{(con)}(t),\boldsymbol{r}(t),\boldsymbol{s}_{i}^{(nor)}(t+1)\}$ from $\mathcal{M}_{i}$.}
  \STATE {Compute $y_{i}=\boldsymbol{r}(t)+\gamma\mathcal{Q}_{i}^{'}\left(\boldsymbol{s}_{i}\left(t+1\right),\chi^{'}_{i}\left(\boldsymbol{s}_{i}(t+1)|\boldsymbol{\varPhi}_{\chi^{'}_{i}}\right)|\boldsymbol{\varPhi}_{\mathcal{Q}^{'}_{i}}\right)$.}
  \STATE {Update Agent ${i}$ critic by minimizing the loss: $\mathcal{L}_{i}^{'}=\frac{1}{\mathcal{B}_{i}} \sum\left(y_{i}-\mathcal{Q}_{i}(\boldsymbol{s}_{i}^{(nor)}(t),\boldsymbol{a}_{i}^{(con)}(t)|\boldsymbol{\varPhi}_{\mathcal{Q}_{i}})\right)^2$. }
  %$\mathcal{L}=\frac{1}{\mathcal{B}} \sum_{\boldsymbol{s}(t),\boldsymbol{a}(t),\boldsymbol{r}(t),\boldsymbol{s}(t+1) \in \mathcal{B}}\left(y-\mathcal{Q}(\boldsymbol{s}(t),\boldsymbol{a}(t)|\boldsymbol{\varPhi}_{\mathcal{Q}})\right)^2$ 
  \STATE {Update Agent ${i}$ actor policy using the sampled policy gradient:
  $\nabla_{\boldsymbol{\varPhi}_{\chi_{i}}}\frac{1}{\mathcal{B}_{i}}\sum_{\boldsymbol{s}_{i} \in \mathcal{B}_{i}}  \mathcal{Q}_{i}(\boldsymbol{s}_{i}^{(nor)}(t),\chi(\boldsymbol{s}^{(nor)}_{i}(t)|\boldsymbol{\varPhi}_{\chi_{i}})|\boldsymbol{\varPhi}_{\mathcal{Q}_{i}})$.}\STATE {Update Agent ${i}$ target networks with $\boldsymbol{\varPhi}_{\chi^{'}_{i}}\leftarrow{\varsigma\boldsymbol{\varPhi}_{\chi_{i}}+(1-\varsigma)\boldsymbol{\varPhi}_{\chi^{'}_{i}}}$ and $\boldsymbol{\varPhi}_{\mathcal{Q}^{'}_{i}}\leftarrow{\varsigma\boldsymbol{\varPhi}_{\mathcal{Q}_{i}}+(1-\varsigma)\boldsymbol{\varPhi}_{\mathcal{Q}^{'}_{i}}}$.}
\ENDFOR
\ENDFOR
\ENDFOR
\end{algorithmic}
\end{algorithm}

Apart from the privacy and security concern, there are mainly two reasons of considering the decentralized RL framework. Firstly, during the training stage, the centralized manner allows the BS to broadcast the sub-actionS to EDs for state update. However, in the distributed manner, the BS firstly receives the EDs state and action to calculate the environment award, then only broadcasts the cooperative reward for distributed training. Secondly, during the execution stage, in the centralized manner, BS still needs to broadcast sub-actions determined by the trained actor network to EDs. Differently, in the distributed manner, EDs can make action decisions based on its own distributed actor networks with their local observation.

Most existing works have focused on homogeneous multi-agent RL with agents sharing the same state and action space. However, in our system, due to the diversity of the action variables, the multi-agent, which includes EDs and BS, have a heterogeneous composition. Different categories of the agents need to leverage their unique information and rely on other agents' specializations to collaboratively find effective policies. %\cite{wakilpoor2020heterogeneous} 
 Therefore, we propose the distributed training and execution framework called HMA-LMIDDPG as illustrated in Fig. \ref{fig:maddpg}(a). Under this framework, both EDs and BS act as the heterogeneous agents to train their independent critic and actor networks. During the execution stage, each ED $u_k$ can take the action decision based on its own local observation without listening to the sub-action instruction from the BS.
Specifically, the heterogeneous multi-agent design is specified. In the proposed distributed RL framework, there are two categories of the heterogeneous multi agents. The first agent category includes all homogeneous EDs:
\begin{itemize}

\item \textbf{State space}: In this paper, when ED acts as the agent, the state $\boldsymbol{s}_{k}(t)$ at each time slot $t$ can be defined as $\boldsymbol{s}_{k}(t)=\{Q_{k}(t),Q_{k}(t-1),\mathcal{P}_{k}(t)\}$. Each ED can only access its own state.
%    with the sub-state at each ED $\boldsymbol{s}_k(t)$ defined as
%    \begin{align}
%        \boldsymbol{s}_k(t)=\{Q_{k}^{(l)}(t)\}.
        %, O^{-}_{k}(t-1),
%    \end{align}
    %\textcolor{red}{Note: the offload decision $\{0,1\}$ and outage fail $\{-1\}$ has been removed, as this has already been reflected in reward for making decision.}
    %where $O^{-}_{k}(t-1)\in \{{O}_{k},-1\}$ denotes the task offloading condition of the previous time slot. Specially, $O^{-}_{k}(t-1)=-1$ represents that it is failed to execute the offloading decision due to the computation task transmission suffers from outage.
    \item \textbf{Action space}: The action vector of the ED $u_k$ can be formulated as $\boldsymbol{a}_k(t)=\{{O}_k^{(e)}(t),{O}_k^{(l)}(t),\\{p}_k(t),{f}_k^{(l)}(t)\}$ with mixed integer property.
\end{itemize}
The second agent category includes the BS, which is heterogeneous compared to EDs:
\begin{itemize}
    \item \textbf{State space}: The state $\boldsymbol{s}_{K+1}(t)$ at each time slot $t$ can be defined as $\boldsymbol{s}_{K+1}(t)=\{R^{(o)}(t),Q(t),\\Q(t-1),\mathcal{P}(t)\}$, with the information acquired from the EDs and BS itself. 
%    with the sub-state at each ED $\boldsymbol{s}_k(t)$ defined as
%    \begin{align}
%        \boldsymbol{s}_k(t)=\{Q_{k}^{(l)}(t)\}.
        %, O^{-}_{k}(t-1),
%    \end{align}
    %\textcolor{red}{Note: the offload decision $\{0,1\}$ and outage fail $\{-1\}$ has been removed, as this has already been reflected in reward for making decision.}
    %where $O^{-}_{k}(t-1)\in \{{O}_{k},-1\}$ denotes the task offloading condition of the previous time slot. Specially, $O^{-}_{k}(t-1)=-1$ represents that it is failed to execute the offloading decision due to the computation task transmission suffers from outage.
    \item \textbf{Action space}: The action vector of the whole system can be formulated as $\boldsymbol{a}_0(t)=\{{f}^{(e)}(t),\boldsymbol{\varTheta}(t)\}$.
\end{itemize} 
What's more, the system reward is defined as the cooperative reward that can be acquired at the BS:
\begin{itemize}
    \item \textbf{Cooperative Reward}: It is similar to the reward function designed in the centralized framework $\boldsymbol{r}(t)=\mathcal{R}(\boldsymbol{s}_{all}(t),\boldsymbol{a}_{all}(t))=-\mathcal{L}(t)$ with  $\boldsymbol{s}_{all}(t)=\{\boldsymbol{s}_{1\rightarrow{K}}(t),\boldsymbol{s}_{K+1}(t)\}$ as the statelist and $\boldsymbol{a}_{all}(t)=\{\boldsymbol{a}_{1\rightarrow{K}}(t),\boldsymbol{a}_{K+1}(t)\}$ as the actionlist.
\end{itemize}

 In our proposed distributed RL framework, during the training stage, the uplink information transmitted between EDs and BS  is the EDs' state space $\boldsymbol{s}_{1\rightarrow{K}}(t)$ and action space $\boldsymbol{a}_{1\rightarrow{K}}(t)$. Moreover, the structure of the actor network for agent EDs and agent BS are illustrated in Fig. \ref{fig:maddpg} (b) and (c), respectively. It can be observed that the defined heterogeneous multi-agent has separate actor network structures from input, activation function, mapping and output. The pseudocode of the training process of the proposed distributed RL algorithm named as HMA-LMIDDPG can be found in \textbf{Algorithm 2}. Although the actor and critic network parameters training process for each agent is similar to the centralized framework in \textbf{Algorithm1}, there are two differences we would like to emphasize. Firstly, as the main purpose of our work is to acquire optimal energy efficiency of the whole system with queue stability, the reward is the cooperative reward which is calculated with the knowledge of the whole system. Therefore, during the training stage, the cooperative Lyapunov-function-based reward is calculated at the BS and then broadcasted to all EDs for training. Secondly, in the designed heterogeneous multi-agent structure, each agent $i\in\{1,...,K+1\}$ has its own replay buffer memory $\mathcal{M}_{i}$ to store their individual experience $\{\boldsymbol{s}_{i}^{(nor)}(t),\boldsymbol{a}_{i}^{(con)}(t),\boldsymbol{r}(t),\boldsymbol{s}_{i}^{(nor)}(t+1)\}$ for later stage training with random mini-batch $\mathcal{B}_{i}$ sampling.

\subsection{Time and Space Complexity}
In this subsection, we would like to discuss the time complexity and the space complexity at both training and execution stages. The time complexity at each agent for individual critic-actor policy training is mainly bounded by $\mathcal{O}\left(\sum_{l=0}^{L_{ac}-1}n_{ac,l}n_{ac,l+1}+\sum_{l^{'}=0}^{L_{cr}-1}n_{cr,l^{'}}n_{cr,l^{'}+1}\right)$, where $L_{ac}$ and $L_{cr}$ are the layer number of the actor and critic network, $n_{ac,l}$ and $n_{cr,l^{'}}$ are the number of neural nodes at each layer of actor and critic network, respectively. During the training stage, apart from the neural weights storage, the experience also needs to be stored in the replay memory buffer for training, thus the space complexity at each agent is bounded by $\mathcal{O}\left(\sum_{l=0}^{L_{ac}-1}n_{ac,l}n_{ac,l+1}+\sum_{l^{'}=0}^{L_{cr}-1}n_{cr,l^{'}}n_{cr,l^{'}+1}\right)+\mathcal{O}(K+M)$. During the execution stage, at each agent, there is only the trained actor network without critic network. Therefore, the complexity of the execution is bounded by $\mathcal{O}\left(\sum_{l=0}^{L_{ac}-1}n_{ac,l}n_{ac,l+1}\right)$. Moreover, the space complexity of the execution is the same $\mathcal{O}\left(\sum_{l=0}^{L_{ac}-1}n_{ac,l}n_{ac,l+1}\right)$.

\section{Numerical Results}
\label{section5}
In this section, the performance of the proposed algorithms is evaluated. We assume that there are mainly $K=4$ mobile EDs at the sub-channel IRS-assisted NOMA-aided MEC system. The coordinates of BS and IRS are $[0,0]$ and $[50,10]$, respectively. The initial locations of all EDs are $\mathcal{P}_{1}(0)=[73,1]$, $\mathcal{P}_{2}(0)=[70,0]$, $\mathcal{P}_{3}(0)=[65,-5]$, $\mathcal{P}_{4}(0)=[70,5]$. The Gauss-Markov mobility related parameters\cite{9687317} of EDs are: $\alpha=[0.4,0.4]$, $\Bar{\kappa}=[2,2]$, and $\Bar{v}=[1,1]$. The distance-dependent path-loss factor equal to $\beta^\text{(DI)}_k=10^{-3}(d^\text{(DI)}_k)^{-2}$, $\beta^\text{(DB)}_k=10^{-3}(d^\text{(DB)}_k)^{-4}$, $\beta^\text{(IB)}=10^{-3}(d^\text{(IB)})^{-2.2}$ with $d^\text{(DI)}_k$, $d^\text{(DB)}_k$, and $d^\text{(IB)}$ are the distances of ED $u_k$ to IRS, ED $u_k$ to BS, and IRS to BS links, respectively. Moreover, the Racian factor $\delta=1$ and the channel noise power is $\sigma^2=-174$dBm\cite{zhi2021ris}. The summary of other system simulation parameters is listed in Table \ref{table2}. The parameters of the critic and actor networks in proposed LMIDDPG and HMA-MIDDPG can also be found in Table \ref{table2}. To show the superior energy efficiency performance of our proposed algorithms, there are several benchmarks we have compared: 
\begin{itemize}
    \item No IRS: there is no IRS in the MEC system,
    \item All Offloading + Random Phase: all the users decide to offload and the phase shifts on the IRS are randomly assigned,
    \item Random Phase: the phase shifts on the IRS are randomly assigned,
    \item LMIA2C: the A2C RL algorithm\cite{mnih2016asynchronous} with Lyapunov-function-based reward and mixed integer mapping.
\end{itemize}
\subsection{Benchmarks Comparison}
\begin{figure*}
\begin{floatrow}
\ffigbox{%
 \includegraphics[width=8.5cm]{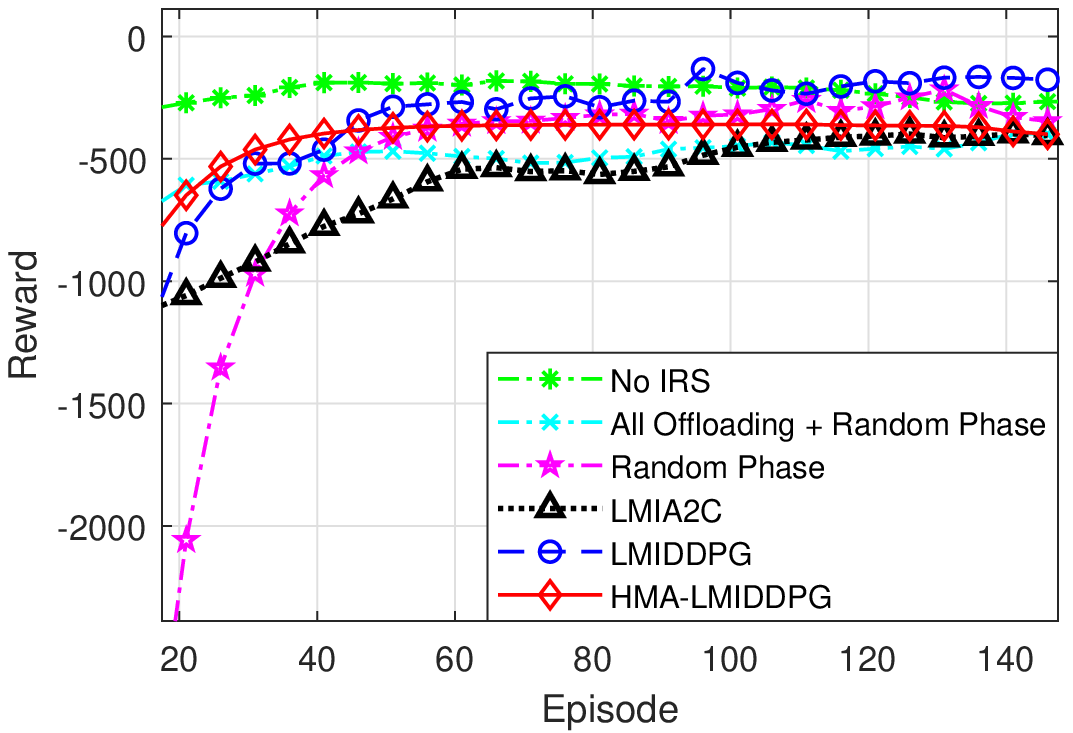}\\%
  
}{%
  \captionof{figure}{The convergence of different algorithms.}\label{fig:fig5}%
}
\ffigbox{%
  \includegraphics[width=8.5cm]{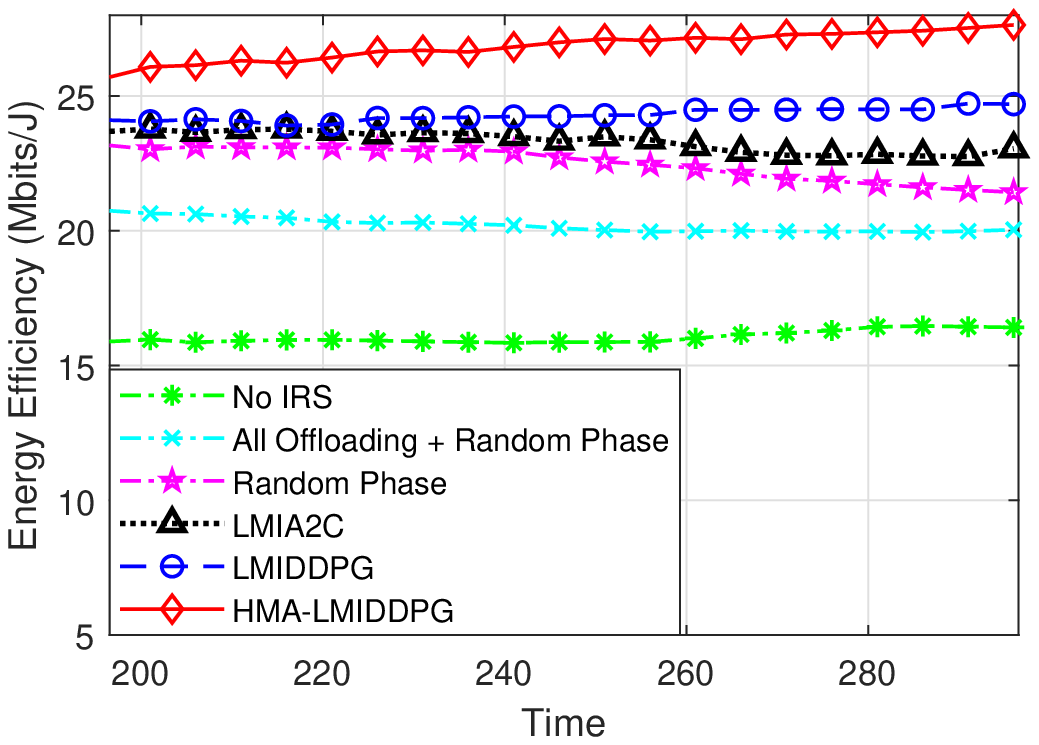}\\%
}{%
  \captionof{figure}{The energy efficiency of different algorithms over time.}\label{fig:fig6}%
}
\end{floatrow}
\end{figure*}
\begin{figure*}
\begin{floatrow}
\capbtabbox{%
  \scalebox{0.65}{
\begin{tabular}{|lc|lc|}
\hline
\multicolumn{2}{|c|}{\textbf{System Parameters}} & \multicolumn{2}{c|}{\textbf{Training Parameters}} \\ \hline
\multicolumn{1}{|c|}{Parameters}     & Value     & \multicolumn{1}{c|}{Parameters}      & Value      \\ \hline
\multicolumn{1}{|c|}{$\Delta T$,  $T$}             & {$1$, $300$}    &    \multicolumn{1}{c|} {$\mathcal{E}$} & {$150$}\\ \hline
\multicolumn{1}{|c|}{$K$}             & {$4$}      &  \multicolumn{1}{c|} {$\gamma$} & {$0.99$}\\ \hline
\multicolumn{1}{|c|}{$M$, $\varpi$\cite{9516969}}             & {$16$, $0.001$W}       & \multicolumn{1}{c|} {$\mathcal{M},\mathcal{M}_{i}$} & {$5000$}\\ \hline
\multicolumn{1}{|c|}{$B^{'}$\cite{9270605}}         & {$2$MHz}        &\multicolumn{1}{c|} {$\mathcal{B}$, $\mathcal{B}_{i}$} & {$128$}\\ \hline
\multicolumn{1}{|c|}{$\zeta_{k}$, $A_{k}$}  &{$0.9$, $3$Mbps}         & \multicolumn{1}{c|} {$\varsigma$} & {$0.01$}\\ \hline
%\multicolumn{1}{|l|}{$A_{k}$}      &{$3$Mbps}     \\ \hline
\multicolumn{1}{|c|}{$\rho_{max}$}      &{$1$W}   & \multicolumn{1}{c|} {$\varrho_{\chi}$} & {$0.0001$} \\ \hline
\multicolumn{1}{|c|}{$f_{k,max}^{(l)},f_{max}^{(e)}$\cite{9467317}} &{$0.3$GHz, $1.26$GHz}&\multicolumn{1}{c|} {$\varrho_{\mathcal{Q}}$} & {$0.0003$}\\ \hline
\multicolumn{1}{|c|}{$\pi^{(l)},\pi^{(e)}$\cite{9467317}} &{$10^{-26}$, $10^{-27}$}&\multicolumn{1}{c|} {$L_{ac}$,$L_{cr}$} & {$5,5$}\\ \hline
\multicolumn{1}{|c|}{$c_{k},c_0$} &{$100,100$ cycles/bit}&\multicolumn{1}{c|} {$n_{ac,1\rightarrow (L_{ac}-1)}$} & {$256$}\\ \hline
%\multicolumn{1}{|l|}{$\varpi$} &{$0.001$W}\\ \hline
\multicolumn{1}{|c|}{$V$} &{$10$}&\multicolumn{1}{c|} {$n_{cr,1\rightarrow (L_{cr}-1)}$} & {$256$}\\ \hline
\end{tabular}}
}{%
  \captionof{table}{Summary of Simulation Parameters}{\label{table2}}%
}

\ffigbox{%
  \includegraphics[width=8.5cm]{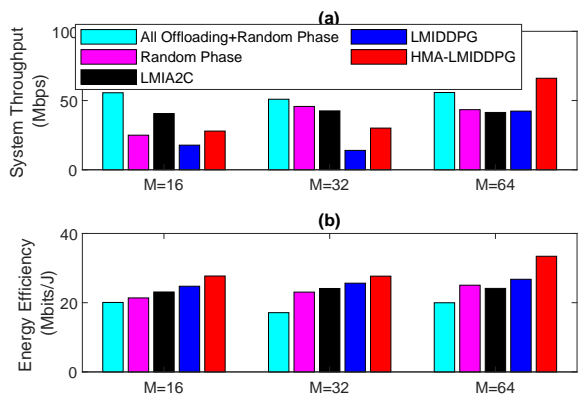}\\%
}{%
  \captionof{figure}{(a) Average system throughput with different IRS sizes $M$, (b) energy efficiency with different IRS sizes $M$ at $T=300$.}  \label{fig:fig7}%
}
\end{floatrow}
\end{figure*}
As shown Fig. \ref{fig:fig5} and Fig. \ref{fig:fig6}, the convergence performance and the energy efficiency have been compared among different benchmarks, respectively. It can be observed that in Fig. \ref{fig:fig5} all the benchmarks and the proposed algorithms can converge at the similar level. However, in Fig. \ref{fig:fig6}, the proposed algorithms LMIDDPG and HMA-LMIDDPG can achieve higher energy efficiency compared to the benchmarks. Specially, compared to No IRS, all the other scenarios and algorithms show higher system energy efficiency, which proves the benefits of deploying IRS in the MEC system. For example, the random phase IRS can achieve around $25\%$ energy efficiency enhancement. The proposed LMIDDPG has around $56\%$ improvement. Furthermore, the proposed distributed framework HMA-HMIDDPG can achieve further $10\%$ improvement compared to centralized framework LMIDDPG.
%In our simulation, the EDs are assumed to follow Gauss-Markov mobility model\cite{7080887}. To be specific, the velocity of $u_k$ at time slot $t$ can be modeled as $V_{k}(t)=\alpha\cdot V_{k}(t-1)+(1-\alpha)\cdot\bar{V}+\bar{\kappa}\cdot\sqrt{1-\alpha^2}\cdot W_{k}(t-1),$ where $V_{k}(t)=\left[v_{k}^{x}(t),v_{k}^{y}(t)\right]$ is the velocity vector, $W_{k}(t)=\left[w_{k}^{x}(t),w_{k}^{y}(t)\right]\sim\mathcal{N}(0,\varsigma^2)$ denotes the uncorrelated random Gaussian process, and $\alpha=[\alpha^{x},\alpha^{y}]$ represents the memory level, $\bar{V}=[\bar{v}^{x},\bar{v}^{y}]$ denotes the asymptotic mean, and $\bar{\kappa}=[\bar{\kappa}^{x},\bar{\kappa}^{y}]$ represents the asymptotic standard deviation of the velocity. Thus, given the velocity of $V_{k}(t)$, the position $\mathcal{P}_{k}(t)=[x_{k}(t),y_{k}(t)]$ at time slot $t$ of ED $u_k$ can be updated as $\mathcal{P}_{k}(t)=\mathcal{P}_{k}(t-1)+V_{k}(t)\Delta T.$
\begin{figure*}
\begin{floatrow}
\ffigbox{%
 \includegraphics[width=8.5cm]{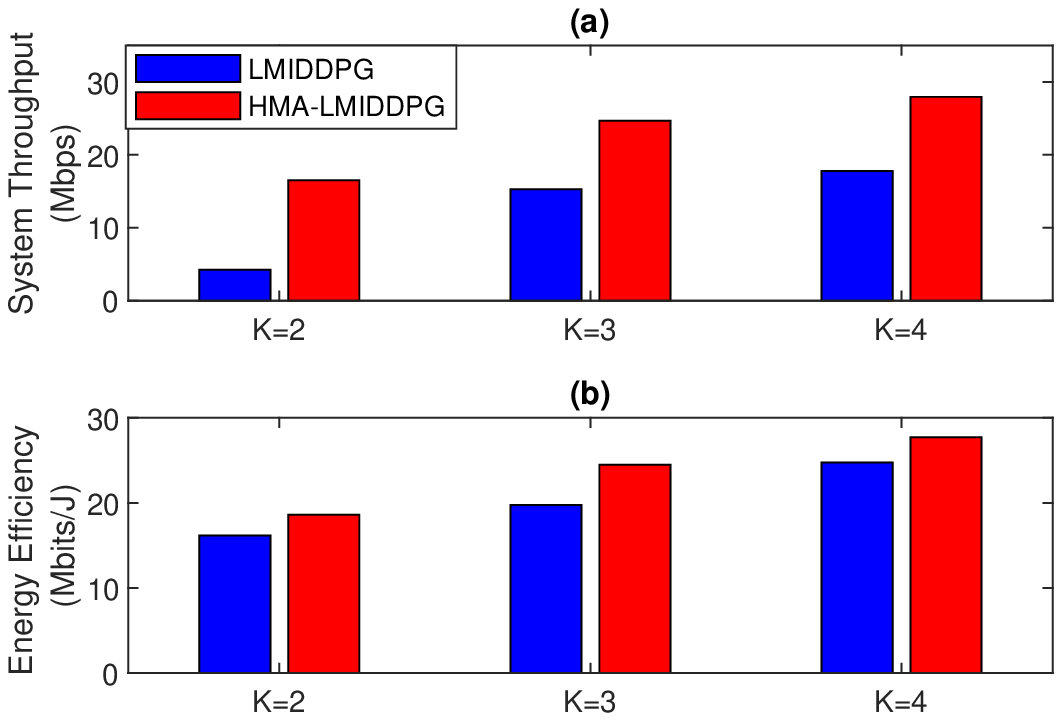}\\%
}{%
  \captionof{figure}{(a) Average system throughput with different user number $K$, (b) energy efficiency with different user number $K$ at $T=300$.}\label{fig:fig8}%
}
\ffigbox{%
  \includegraphics[width=8.5cm]{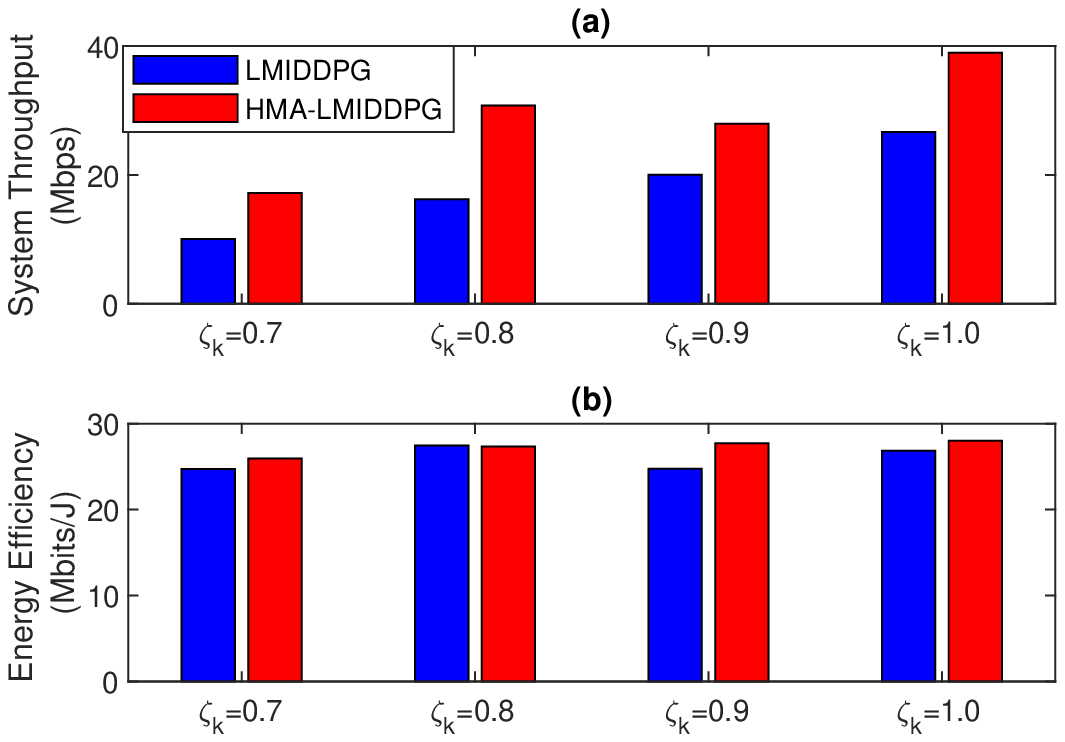}\\%
}{%
  \captionof{figure}{(a) Average system throughput with different task generation probability $\zeta_{k}$, (b) energy efficiency with different task generation probability $\zeta_{k}$ at $T=300$.}  \label{fig:fig9}%
}
\end{floatrow}
\end{figure*}

In Fig. \ref{fig:fig7}, we further compare the average system throughput over $T=300$ and the energy efficiency at $T=300$ based on different IRS element sizes $M$. It can be observed in Fig. \ref{fig:fig7} (a), the All Offloading scenario always achieves higher system throughput due to the constant task offloading transmission. However, the energy efficiency of which is very low as shown in Fig. \ref{fig:fig7} (b). Differently, the proposed algorithms, LMIDDPG and HMA-MIDDPG, can acquire the highest energy efficiency, which proves the superior resource allocation decision. Moreover, in Fig. \ref{fig:fig7} (a), apart from all offloading benchmark, it can be observed that the lower throughput proves the different offloading and local computing decisions have been made. What's more, with lower system throughput, the energy consumption is much lower to achieve higher energy efficiency performance. 
\subsection{Impact of Different Parameters}
In Fig. \ref{fig:fig8}, the average system throughput over $T=300$ and the energy efficiency at $T=300$ based on different user number $K$ have been compared among the proposed centralized framework LMIDDPG and distributed framework HMA-LMIDDPG. As shown in Fig. \ref{fig:fig8} (a), with the user number increasing, the overall throughput increases accordingly under both frameworks. Moreover, from Fig. \ref{fig:fig8} (b), HMA-LMIDDPG can achieve higher system throughput with higher energy efficiency, which proves the superior decision making based on the proposed distributed framework. 

In Fig. \ref{fig:fig9}, the average system throughput over $T=300$ and the energy efficiency at $T=300$ based on different task generation probability $\zeta_{k}$ have been compared among the proposed centralized framework LMIDDPG and distributed framework HMA-LMIDDPG. It is easily noticed in Fig. \ref{fig:fig9} (a) that with higher task generation probability, the system throughput becomes higher. However, Fig. \ref{fig:fig9} (b) shows that the energy efficiency has similar performance among different probabilities.

\begin{figure*}
\begin{floatrow}
\ffigbox{%
 \includegraphics[width=8.5cm]{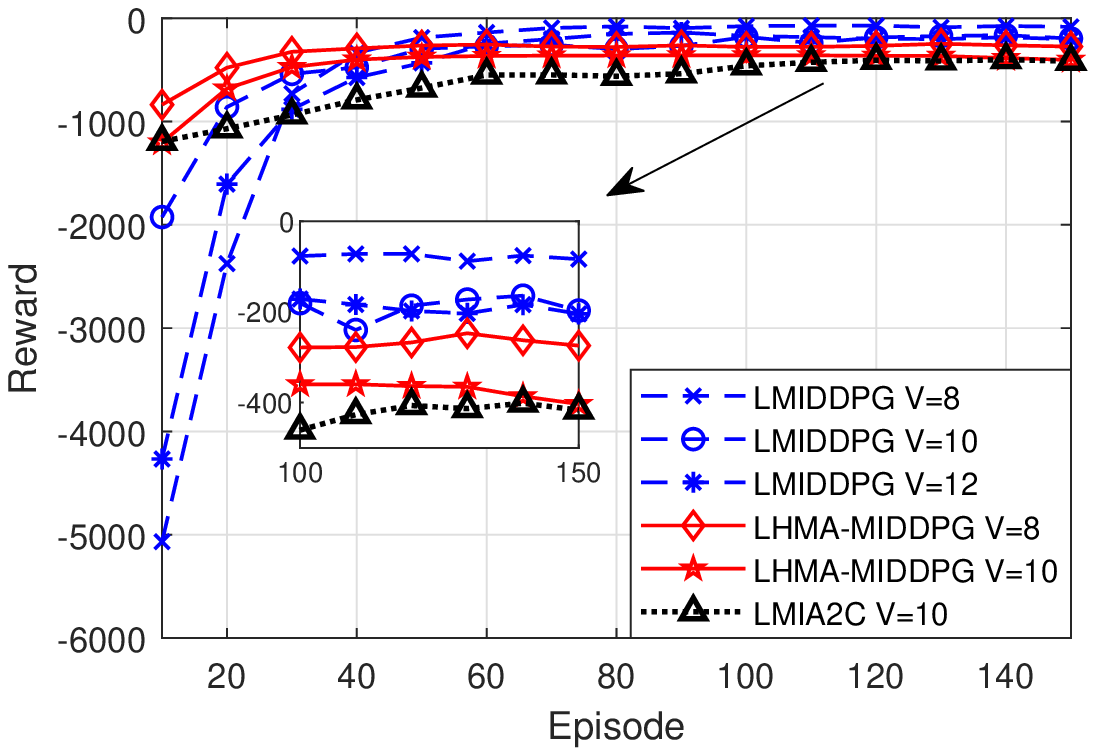}\\%
}{%
  \captionof{figure}{The convergence of different algorithms with different Lyapunov weight $V$.}\label{fig:fig10}%
}
\ffigbox{%
  \includegraphics[width=8.5cm]{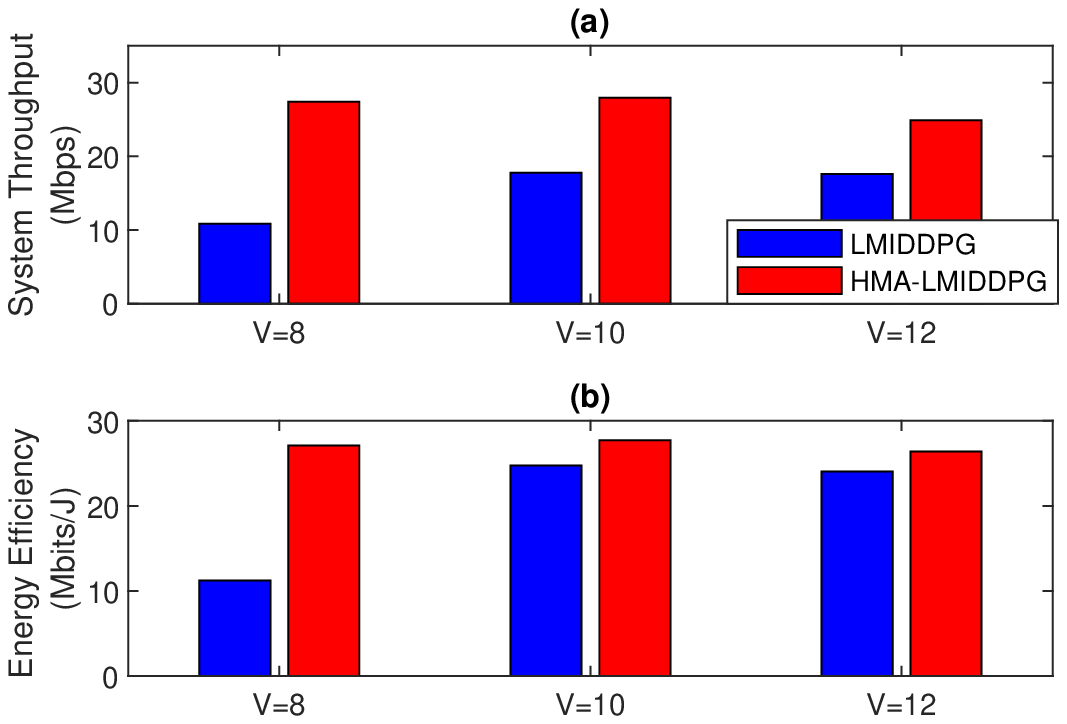}\\%
}{%
  \captionof{figure}{(a) Average system throughput with different Lyapunov weight $V$, (b) energy efficiency with different Lyapunov weight $V$ at $T=300$.} \label{fig:fig11}%
}
\end{floatrow}
\end{figure*}

As shown in Fig. \ref{fig:fig10} and Fig. \ref{fig:fig11}, the convergence performance, the average system throughput over $T=300$ and the energy efficiency at $T=300$ based on different Lyapunov weights $V$ have been illustrated, respectively. In Fig. \ref{fig:fig10}, compared to LMIA2C, the proposed algorithms can converge at higher rewards. Moreover, with $V$ increases the rewards decreases. This trend is because the design of the reward function is the negative eq. (\ref{driftpluspenalty}) with negative $V$. Moreover, from Fig. \ref{fig:fig11} (a) that with $V$ increases, the system throughput also increases under centralized framework LMIDDPG. From Fig. \ref{fig:fig11} (b), the proposed distributed framework HMA-LMIDDPG always achieves better energy efficiency performance.
\begin{figure*}
\begin{floatrow}
\ffigbox{%
 \includegraphics[width=8.5cm]{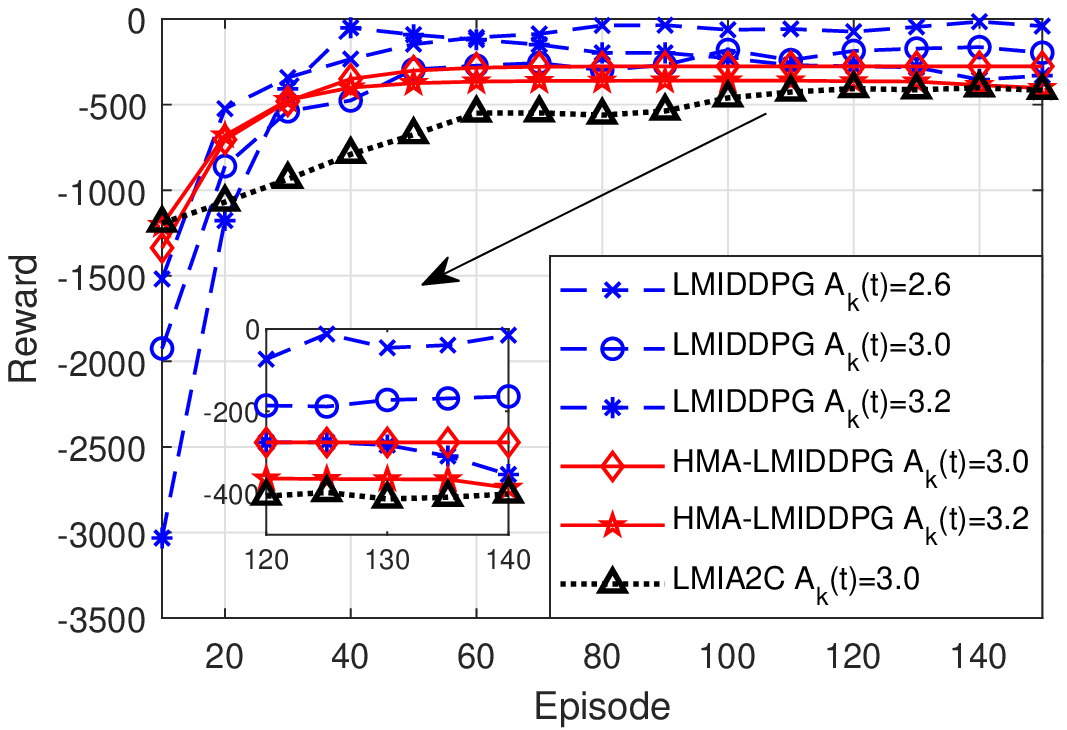}\\%
}{%
  \captionof{figure}{The convergence of different algorithms with different arrival data sizes $A_{k}(t)$.}\label{fig:fig12}%
}
\ffigbox{%
  \includegraphics[width=8.5cm]{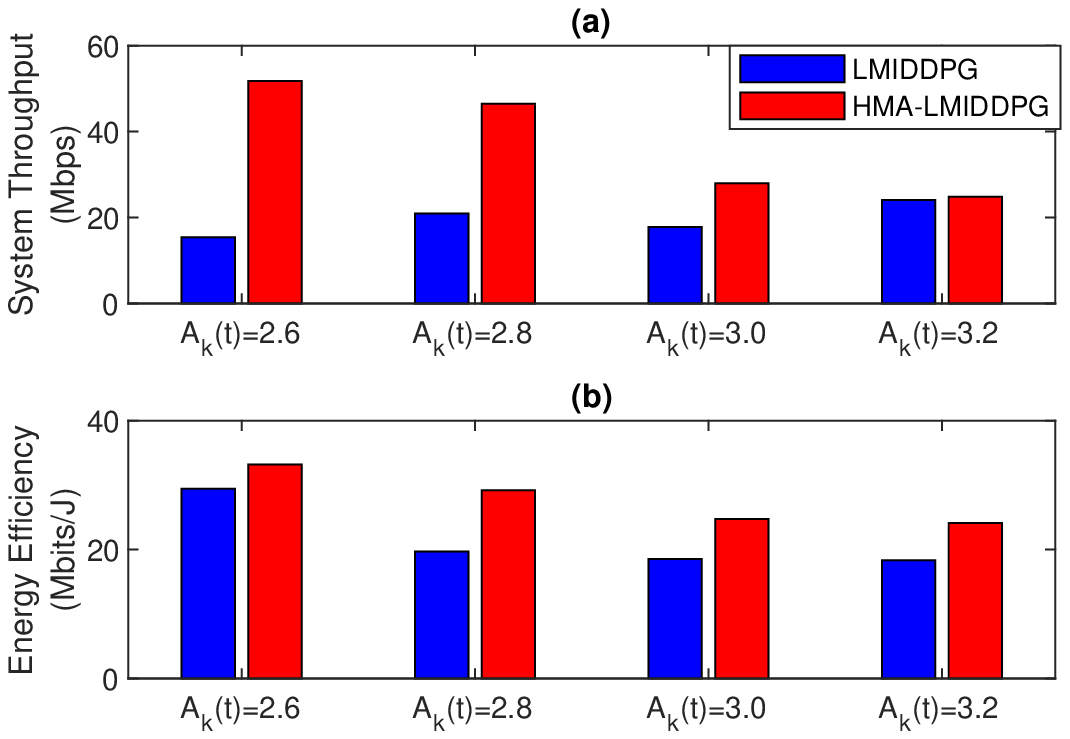}\\%
}{%
  \captionof{figure}{(a) Average system throughput with different arrival data sizes $A_{k}(t)$, (b) energy efficiency with different arrival data sizes $A_{k}(t)$ at $T=300$.}\label{fig:fig13}%
}
\end{floatrow}
\end{figure*}
\begin{figure*}
\begin{floatrow}
\ffigbox{%
 \includegraphics[width=8.5cm]{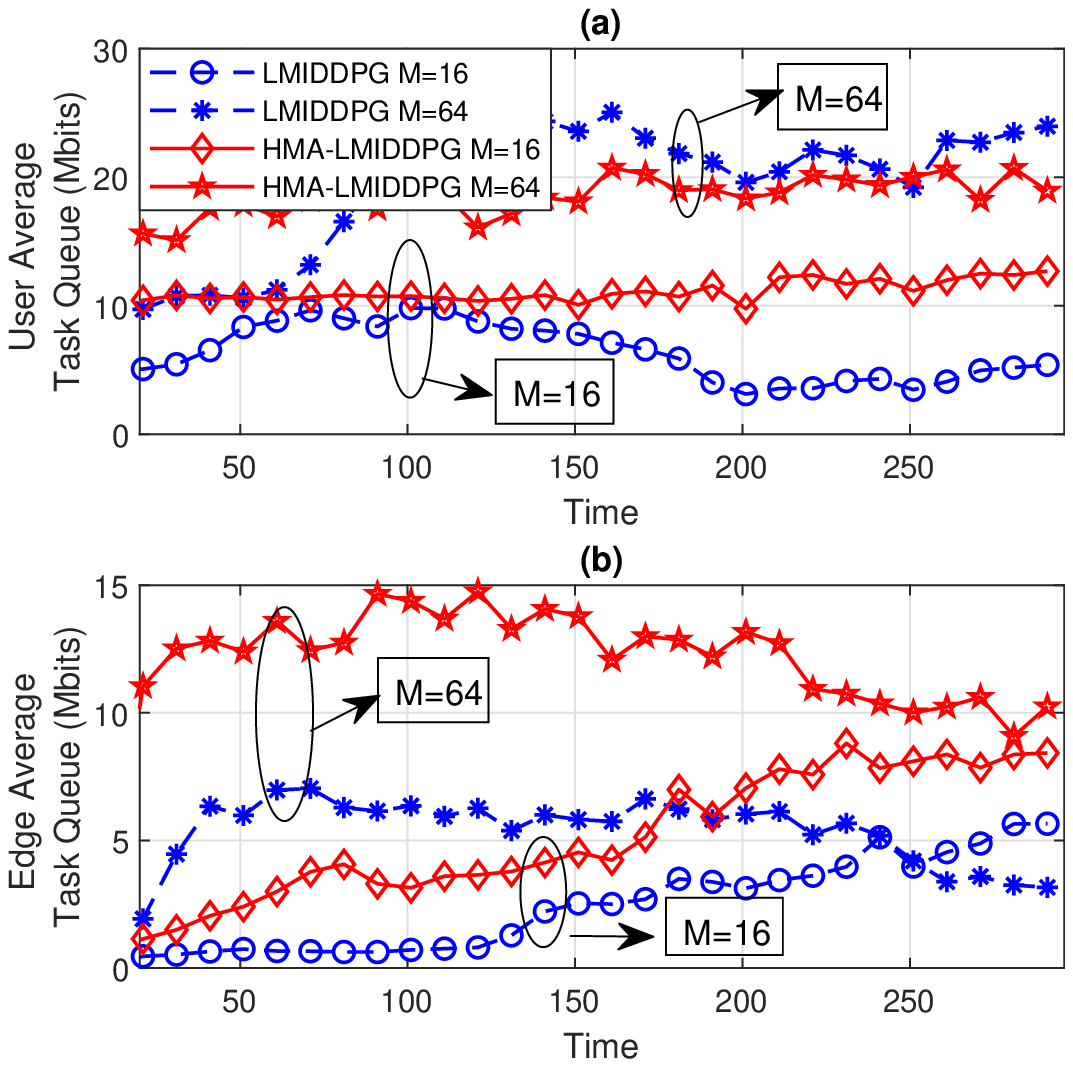}\\%
}{%
  \captionof{figure}{The queue stability with different IRS sizes $M$. (a) User average task queue over time, (b) edge average task queue over time.}\label{fig:fig15}%
}
\ffigbox{%
  \includegraphics[width=8.5cm]{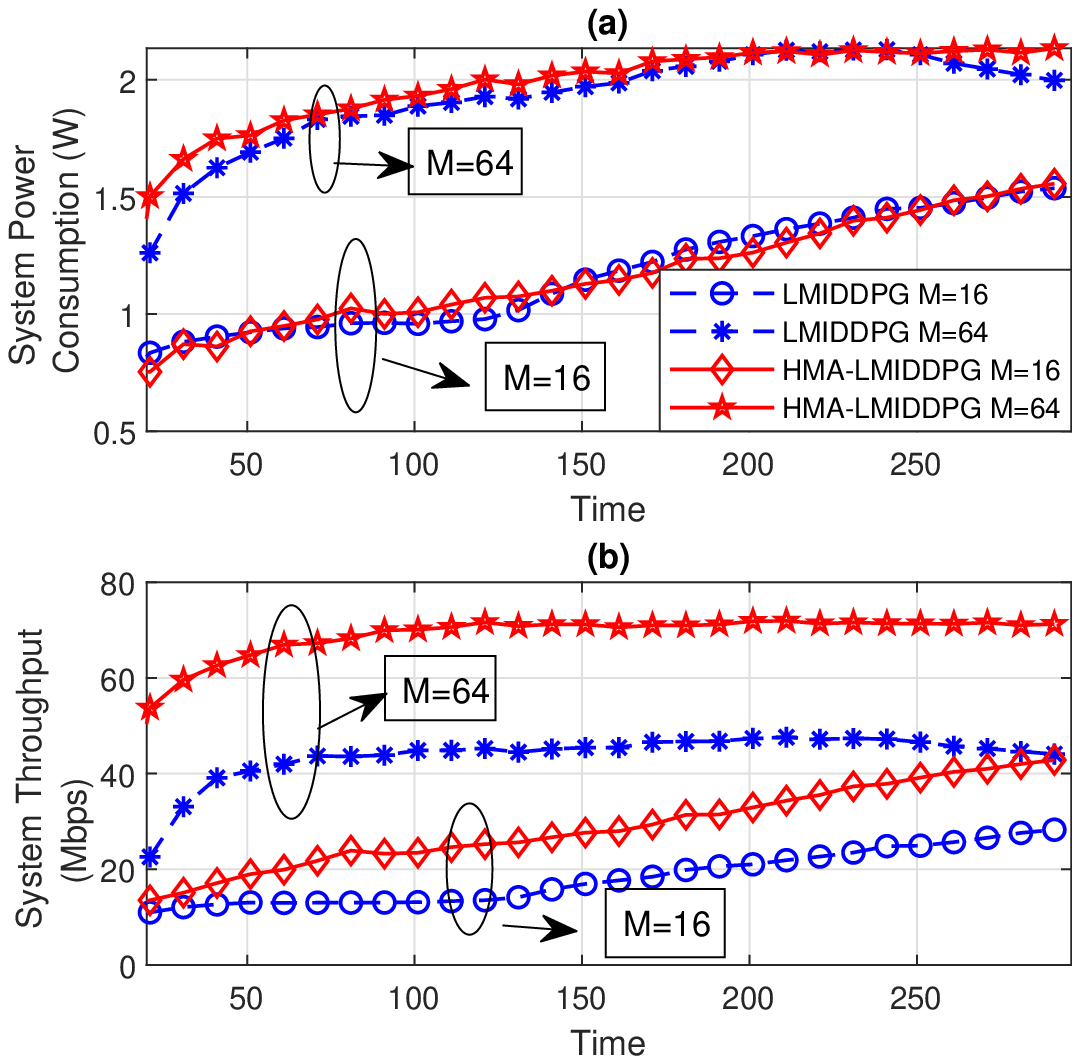}\\%
}{%
  \captionof{figure}{(a) System power consumption over time, (b) system throughput over time.}  \label{fig:fig14}%
}
\end{floatrow}
\end{figure*}

In Fig. \ref{fig:fig12} and Fig. \ref{fig:fig13}, the convergence performance, the average system throughput over $T=300$ and the energy efficiency at $T=300$ based on different arrival data sizes $A_{k}(t)$ are plotted, respectively. It is easily noticed in Fig. \ref{fig:fig12} that with $A_{k}(t)$ increases, the rewards decreases. This is because of the design of the reward function is the negative eq. (\ref{driftpluspenalty}) with upper bounds obtained from eq. (\ref{upper}) with arrival data. Moreover, as shown in Fig. \ref{fig:fig13}, with arrival data size increases, the distributed framework HMA-LMIDDPG can always achieve higher average system throughput and higher energy efficiency compared to LMIDDPG. 
\subsection{System Performance}
Fig. \ref{fig:fig15} (a) and Fig. \ref{fig:fig15} (b) show the average stable queue status at both user EDs and edge BS, separately. What's more, with the knowledge of the superiority of our proposed algorithms, in this subsection, we evaluate the detailed performance of the system over time. In Fig. \ref{fig:fig14}, the performance of the system power consumption and throughput over time are evaluated based on different IRS element sizes $M=16$ and $M=64$. Based on the conclusion that HMA-LMIDDPG can always achieve high energy efficiency compared to LMIDDPG, interestingly, for both $M=16$ and $M=64$, we can observed that the proposed algorithms have similar system power consumption from Fig. \ref{fig:fig14} (a). However, from Fig. \ref{fig:fig14} (b), compared to LMIDDPG, the distributed HMA-LMIDDPG maintains higher system throughput with time evolving. 
\section{Conclusions}
\label{section6}
In this paper, we investigated the joint offloading, communication and computation resource allocation for IRS-assisted NOMA-aided MEC system. We proposed the mixed integer energy efficiency maximization problem with system queue stability constraint. The centralized RL framework algorithm called LMIDDPG has been proposed with the award  function defined as the upper-bound of the Lyapunov drift-plus-penalty function. In addition, we further designed the mixed integer action space mapping which contains both continuous mapping and integer mapping. To ensure less communication interaction overheads between BS and EDs during the execution stage, HMA-LMIDDPG has been further proposed with heterogeneous multi-agent that includes both homogeneous agents EDs and heterogeneous agent BS. Under this framework, each agent can make its individual action decisions based on its local observation. Numerical results clearly proved that, our proposed algorithms can greatly increase the system energy efficiency with maintained queue stability compared to benchmark algorithms. Moreover, the distributed structure HMA-LMIDDPG can achieve more energy efficiency gain compared to centralized structure LMIDDPG.  

\appendices

\section{Lyapunov Drift-plus-penalty Upper Bound}
The drift item in eq.(\ref{driftpluspenalty}) can be extended as
\begin{equation}
\begin{split}
&\Delta L\left(\boldsymbol{Q}(t)\right)=
\frac{1}{2}\left(\boldsymbol{Q}^{2}(t+1)-\boldsymbol{Q}^{2}(t)\right)=
    \frac{1}{2}\left(\left(\boldsymbol{Q}(t)+\boldsymbol{Q}(t)^{(in)}-\boldsymbol{Q}(t)^{(out)}\right)^2 -\boldsymbol{Q}^{2}(t)\right)
    \\&=\frac{1}{2} \left(\boldsymbol{Q}(t)^{(in)}-\boldsymbol{Q}(t)^{(out)}\right)\left(2\boldsymbol{Q}(t)+\boldsymbol{Q}(t)^{(in)}-\boldsymbol{Q}(t)^{(out)}\right)
    \\&=\frac{1}{2}\left(\boldsymbol{Q}(t)^{(in)}-\boldsymbol{Q}(t)^{(out)}\right)^{2}+\boldsymbol{Q}(t)\left(\boldsymbol{Q}(t)^{(in)}-\boldsymbol{Q}(t)^{(out)}\right)
    \\&\leqslant \mathbb{E}\left\{\sum_{j=1}^{K+1}\frac{1}{2}\left(Q_{j}^{(in)}(t)^2+Q_{j}^{(out)}(t)^2\right)|\boldsymbol{Q}(t)\right\}+\mathbb{E}\left\{\sum_{j=1}^{K+1}Q_{j}(t)\left(Q_{j}^{(in)}(t)-Q_{j}^{(out)}(t)\right)|\boldsymbol{Q}(t)\right\}
    \\&\leqslant b+\mathbb{E}\left\{\sum_{j=1}^{K+1}Q_{j}(t)\left(Q_{j}^{(in)}(t)-Q_{j}^{(out)}(t)\right)|\boldsymbol{Q}(t)\right\}.
    \end{split}
\end{equation}
Specifically, $b$ is a constant upper bounds which can be obtained as
\begin{equation}
    \begin{split}
        &\mathbb{E}\left\{\sum_{j=1}^{K+1}\frac{1}{2}\left(Q_{j}^{(in)}(t)^2+Q_{j}^{(out)}(t)^2\right)|\boldsymbol{Q}(t)\right\}\leqslant\frac{1}{2}\sum_{j=1}^{K}\left(A_{i}(t)\Delta T\right)^2\\&+\frac{1}{2}\sum_{j=1}^{K}\left(\varLambda_{k,max}^{(l)}(t)\Delta T+R_{k,max}^{(o)}(t)\Delta T\right)^2
        +\frac{1}{2}\left(R_{max}^{(o)}(t)\Delta T \right)^2+ \frac{1}{2}\left(\varLambda^{(e)}_{max}(t)\Delta T\right)^2\triangleq b,
    \end{split}
    \label{upper}
\end{equation}
where the first two items in the inequality denotes the maximum input and the processed (i.e. local computing and offloaded) data size at $K$ EDs task queues, the last two items represents the maximum arrived and the processed data size in the edge task queue at the BS.

\bibliographystyle{IEEEtran}
\bibliography{IEEEabrv,TWC}

% Generated by IEEEtran.bst, version: 1.14 (2015/08/26)
\begin{thebibliography}{10}
\providecommand{\url}[1]{#1}
\csname url@samestyle\endcsname
\providecommand{\newblock}{\relax}
\providecommand{\bibinfo}[2]{#2}
\providecommand{\BIBentrySTDinterwordspacing}{\spaceskip=0pt\relax}
\providecommand{\BIBentryALTinterwordstretchfactor}{4}
\providecommand{\BIBentryALTinterwordspacing}{\spaceskip=\fontdimen2\font plus
\BIBentryALTinterwordstretchfactor\fontdimen3\font minus
  \fontdimen4\font\relax}
\providecommand{\BIBforeignlanguage}[2]{{%
\expandafter\ifx\csname l@#1\endcsname\relax
\typeout{** WARNING: IEEEtran.bst: No hyphenation pattern has been}%
\typeout{** loaded for the language `#1'. Using the pattern for}%
\typeout{** the default language instead.}%
\else
\language=\csname l@#1\endcsname
\fi
#2}}
\providecommand{\BIBdecl}{\relax}
\BIBdecl

\bibitem{7123563}
A.~Al-Fuqaha, M.~Guizani \emph{et~al.}, ``Internet of things: A survey on
  enabling technologies, protocols, and applications,'' \emph{IEEE Commun.
  Surv. Tut.}, vol.~17, no.~4, pp. 2347--2376, 2015.

\bibitem{7488250}
W.~Shi, J.~Cao, Q.~Zhang, Y.~Li, and L.~Xu, ``Edge computing: Vision and
  challenges,'' \emph{IEEE Internet Things J.}, vol.~3, no.~5, pp. 637--646,
  2016.

\bibitem{9113305}
Q.-V. Pham, F.~Fang \emph{et~al.}, ``A survey of multi-access edge computing in
  {5G} and beyond: Fundamentals, technology integration, and
  state-of-the-art,'' \emph{IEEE Access}, vol.~8, pp. 116\,974--117\,017, 2020.

\bibitem{9240028}
J.~Zhu \emph{et~al.}, ``Power efficient {IRS}-assisted {NOMA},'' \emph{IEEE
  Trans. Commun.}, vol.~69, no.~2, pp. 900--913, 2021.

\bibitem{9241881}
A.~S.~d. Sena \emph{et~al.}, ``What role do intelligent reflecting surfaces
  play in multi-antenna non-orthogonal multiple access?'' \emph{IEEE Wireless
  Commun.}, vol.~27, no.~5, pp. 24--31, 2020.

\bibitem{8016573}
Y.~Mao, C.~You, J.~Zhang, K.~Huang, and K.~B. Letaief, ``A survey on mobile
  edge computing: The communication perspective,'' \emph{IEEE Commun. Surv.
  Tut.}, vol.~19, no.~4, pp. 2322--2358, 2017.

\bibitem{8972358}
X.~Liu, J.~Yu, J.~Wang, and Y.~Gao, ``Resource allocation with edge computing
  in {IoT} networks via machine learning,'' \emph{IEEE Internet Things J.},
  vol.~7, no.~4, pp. 3415--3426, 2020.

\bibitem{8598893}
M.~Min, L.~Xiao, Y.~Chen, P.~Cheng, D.~Wu, and W.~Zhuang, ``Learning-based
  computation offloading for {IoT} devices with energy harvesting,'' \emph{IEEE
  Trans. Veh. Technol.}, vol.~68, no.~2, pp. 1930--1941, 2019.

\bibitem{8493155}
X.~Chen, H.~Zhang, C.~Wu, S.~Mao, Y.~Ji, and M.~Bennis, ``Optimized computation
  offloading performance in virtual edge computing systems via deep
  reinforcement learning,'' \emph{IEEE Internet Things J.}, vol.~6, no.~3, pp.
  4005--4018, 2019.

\bibitem{9205989}
X.~Liu, J.~Yu, Z.~Feng, and Y.~Gao, ``Multi-agent reinforcement learning for
  resource allocation in {IoT} networks with edge computing,'' \emph{China
  Communications}, vol.~17, no.~9, pp. 220--236, 2020.

\bibitem{9485089}
X.~Huang, S.~Leng, S.~Maharjan, and Y.~Zhang, ``Multi-agent deep reinforcement
  learning for computation offloading and interference coordination in small
  cell networks,'' \emph{IEEE Trans. Veh. Technol.}, vol.~70, no.~9, pp.
  9282--9293, 2021.

\bibitem{9372298}
A.~Feriani and E.~Hossain, ``Single and multi-agent deep reinforcement learning
  for {AI}-enabled wireless networks: A tutorial,'' \emph{IEEE Commun. Surveys
  Tuts.}, vol.~23, no.~2, pp. 1226--1252, 2021.

\bibitem{7973146}
Z.~Ding, X.~Lei \emph{et~al.}, ``A survey on non-orthogonal multiple access for
  {5G} networks: Research challenges and future trends,'' \emph{IEEE J. Sel.
  Areas Commun.}, vol.~35, no.~10, pp. 2181--2195, 2017.

\bibitem{9679390}
Z.~Ding \emph{et~al.}, ``Hybrid {NOMA} offloading in multi-user {MEC}
  networks,'' \emph{IEEE Trans. Wireless Commun.}, pp. 1--1, 2022.

\bibitem{9393794}
L.~Liu, B.~Sun, X.~Tan, and D.~H.~K. Tsang, ``Energy-efficient resource
  allocation and subchannel assignment for {NOMA}-enabled multiaccess edge
  computing,'' \emph{IEEE Systems Journal}, vol.~16, no.~1, pp. 1558--1569,
  2022.

\bibitem{8794550}
M.~Zeng, N.-P. Nguyen, O.~A. Dobre, and H.~V. Poor, ``Delay minimization for
  {NOMA}-assisted {MEC} under power and energy constraints,'' \emph{IEEE
  Wireless Commun. Lett.}, vol.~8, no.~6, pp. 1657--1661, 2019.

\bibitem{9340353}
L.~Liu, B.~Sun, Y.~Wu, and D.~H.~K. Tsang, ``Latency optimization for
  computation offloading with hybrid {NOMA–OMA} transmission,'' \emph{IEEE
  Internet Things J.}, vol.~8, no.~8, pp. 6677--6691, 2021.

\bibitem{9113721}
L.~Qian, Y.~Wu \emph{et~al.}, ``{NOMA} assisted multi-task multi-access mobile
  edge computing via deep reinforcement learning for industrial internet of
  things,'' \emph{IEEE Trans. Ind. Informat.}, vol.~17, no.~8, pp. 5688--5698,
  2021.

\bibitem{9467317}
Z.~Chen, L.~Zhang \emph{et~al.}, ``{NOMA}-based multi-user mobile edge
  computation offloading via cooperative multi-agent deep reinforcement
  learning,'' \emph{IEEE Trans. Cognit. Commun. Netw.}, vol.~8, no.~1, pp.
  350--364, 2022.

\bibitem{8910627}
Q.~Wu and R.~Zhang, ``Towards smart and reconfigurable environment: Intelligent
  reflecting surface aided wireless network,'' \emph{IEEE Commun. Mag.},
  vol.~58, no.~1, pp. 106--112, 2020.

\bibitem{9388935}
T.~Bai, C.~Pan \emph{et~al.}, ``Resource allocation for intelligent reflecting
  surface aided wireless powered mobile edge computing in {OFDM} systems,''
  \emph{IEEE Trans. Wireless Commun.}, vol.~20, no.~8, pp. 5389--5407, 2021.

\bibitem{9270605}
Z.~Chu, P.~Xiao, M.~Shojafar, D.~Mi, J.~Mao, and W.~Hao, ``Intelligent
  reflecting surface assisted mobile edge computing for internet of things,''
  \emph{IEEE Wireless Commun. Lett.}, vol.~10, no.~3, pp. 619--623, 2021.

\bibitem{9516969}
S.~Mao, N.~Zhang \emph{et~al.}, ``Computation rate maximization for intelligent
  reflecting surface enhanced wireless powered mobile edge computing
  networks,'' \emph{IEEE Trans. Veh. Technol.}, vol.~70, no.~10, pp.
  10\,820--10\,831, 2021.

\bibitem{9449944}
S.~Bi, L.~Huang, H.~Wang, and Y.-J.~A. Zhang, ``Lyapunov-guided deep
  reinforcement learning for stable online computation offloading in
  mobile-edge computing networks,'' \emph{IEEE Trans. Wireless Commun.},
  vol.~20, no.~11, pp. 7519--7537, 2021.

\bibitem{7080887}
S.~Batabyal and P.~Bhaumik, ``Mobility models, traces and impact of mobility on
  opportunistic routing algorithms: A survey,'' \emph{IEEE Commun. Surveys
  Tuts.}, vol.~17, no.~3, pp. 1679--1707, 2015.

\bibitem{yang2020intelligent}
Y.~Yang, B.~Zheng, S.~Zhang, and R.~Zhang, ``{Intelligent reflecting surface
  meets OFDM: protocol design and rate maximization},'' \emph{{IEEE} Trans.
  Wireless Commun.}, vol.~68, no.~7, pp. 4522--4535, Mar 2020.

\bibitem{8904347}
M.~Hua \emph{et~al.}, ``Energy efficient task offloading in {NOMA}-based mobile
  edge computing system,'' in \emph{2019 IEEE 30th Annual International
  Symposium on Personal, Indoor and Mobile Radio Communications (PIMRC)}, Sep
  2019, pp. 1--7.

\bibitem{9214497}
B.~Lyu, P.~Ramezani, D.~T. Hoang, S.~Gong, Z.~Yang, and A.~Jamalipour,
  ``Optimized energy and information relaying in self-sustainable
  {IRS}-empowered {WPCN},'' \emph{{IEEE} Trans. Wireless Commun.}, vol.~69,
  no.~1, pp. 619--633, Oct 2021.

\bibitem{neely2010stochastic}
M.~J. Neely, ``Stochastic network optimization with application to
  communication and queueing systems,'' \emph{Synthesis Lectures on
  Communication Networks}, vol.~3, no.~1, pp. 1--211, 2010.

\bibitem{6893054}
Y.~Li, M.~Sheng, Y.~Shi, X.~Ma, and W.~Jiao, ``Energy efficiency and delay
  tradeoff for time-varying and interference-free wireless networks,''
  \emph{{IEEE} Trans. Wireless Commun.}, vol.~13, no.~11, pp. 5921--5931, Sep
  2014.

\bibitem{9152999}
L.~Bracciale and P.~Loreti, ``Lyapunov drift-plus-penalty optimization for
  queues with finite capacity,'' \emph{IEEE Commun. Lett.}, vol.~24, no.~11,
  pp. 2555--2558, Jul 2020.

\bibitem{schulman2017proximal}
J.~Schulman \emph{et~al.}, ``Proximal policy optimization algorithms,''
  \emph{arXiv preprint arXiv:1707.06347}, 2017.

\bibitem{mnih2016asynchronous}
V.~Mnih, A.~P. Badia, M.~Mirza, A.~Graves, T.~Lillicrap, T.~Harley, D.~Silver,
  and K.~Kavukcuoglu, ``Asynchronous methods for deep reinforcement learning,''
  in \emph{International conference on machine learning}.\hskip 1em plus 0.5em
  minus 0.4em\relax PMLR, 2016, pp. 1928--1937.

\bibitem{lillicrap2015continuous}
T.~P. Lillicrap, J.~J. Hunt, A.~Pritzel, N.~Heess, T.~Erez, Y.~Tassa,
  D.~Silver, and D.~Wierstra, ``Continuous control with deep reinforcement
  learning,'' \emph{arXiv preprint arXiv:1509.02971}, Sep. 2015.

\bibitem{9687317}
Z.~Yang, S.~Bi, and Y.-J.~A. Zhang, ``{Online trajectory and resource
  optimization for stochastic UAV-enabled MEC system},'' \emph{IEEE Trans.
  Wireless Commun.}, pp. 1--1, 2022.

\bibitem{zhi2021ris}
K.~Zhi, C.~Pan, G.~Zhou, H.~Ren, M.~Elkashlan, and R.~Schober, ``{Is RIS-aided
  massive MIMO promising with ZF detectors and imperfect CSI?}'' \emph{arXiv
  preprint arXiv:2111.01585}, Nov. 2021.

\end{thebibliography}
\end{document}